\documentclass[a4paper]{article}

\pdfoutput=1
\usepackage[utf8]{inputenc}

\usepackage[super,sort&compress,comma]{natbib} 
\newcommand{\supercite}[1]{\cite{#1}}

\usepackage[english]{babel}
\usepackage[T1]{fontenc}
\usepackage{csquotes}

\usepackage[a4paper,top=3cm,bottom=2cm,left=3cm,right=3cm,marginparwidth=1.75cm]{geometry}

\usepackage{amsmath}
\usepackage{graphicx}
\usepackage{hyperref}
\hypersetup{colorlinks=true, allcolors=blue}
\usepackage{authblk}
\usepackage{multirow}

\usepackage{nomencl}
\usepackage{booktabs}
\usepackage{tabularx}
\usepackage{textcomp} 
\usepackage{amsfonts}
\usepackage{outlines} 
\usepackage{subfig}
\usepackage[version=4]{mhchem}
\usepackage{siunitx}
\usepackage{bm}
\usepackage[normalem]{ulem} 
\date{}

\DeclareSIUnit\angstrom{\text {Å}}

\usepackage[final]{changes}

\title{Neural networks trained on synthetically generated crystals can extract structural information from ICSD powder X-ray diffractograms}

\author[1,2]{Henrik Schopmans}
\author[1,2]{Patrick Reiser}
\author[1,2]{Pascal Friederich*}

\affil[1]{Institute of Theoretical Informatics, Karlsruhe Institute of Technology, Engler-Bunte-Ring 8, 76131 Karlsruhe, Germany}
\affil[2]{Institute of Nanotechnology, Karlsruhe Institute of Technology, Hermann-von-Helmholtz-Platz 1, 76344 Eggenstein-Leopoldshafen, Germany}
\affil[*]{Corresponding author: pascal.friederich@kit.edu}

\begin{document}

\maketitle
\begin{abstract}
    Machine learning techniques have successfully been used to extract structural information such as the crystal space group from powder X-ray diffractograms.
However, training directly on simulated diffractograms from databases such as
the ICSD is challenging due to its limited size, class-inhomogeneity, and bias
toward certain structure types. We propose an alternative approach of generating
synthetic crystals with random coordinates by using the symmetry operations of
each space group. Based on this approach, we 
demonstrate online training of deep ResNet-like models on up to a few
million unique on-the-fly generated synthetic diffractograms per hour. For our chosen task of
space group classification, we achieved a test accuracy of 79.9\% on unseen ICSD
structure types from most space groups. This surpasses the 56.1\% accuracy of the current state-of-the-art approach of training on ICSD crystals directly. Our results demonstrate that synthetically generated crystals can be
used to extract structural information from ICSD powder diffractograms, which makes it possible to apply very large state-of-the-art machine learning models in the area of powder X-ray diffraction.
We further show first steps toward applying our methodology to experimental data, where automated XRD data analysis is crucial, especially in high-throughput settings.
While we focused on the prediction of the space group, our approach has the potential to be extended to related tasks in the future.
\end{abstract}

\section{Introduction}\label{sec:introduction}
Machine learning techniques have emerged as a powerful tool in the toolkit of
materials scientists. While they are often used to make predictions on
the properties of materials or find new materials with certain properties, an increasingly interesting domain is the automated analysis of raw experimental measurements
guided by machine learning\supercite{radovicMachineLearningEnergy2018}.

With the advent of high-throughput experiments, the amount of gathered data is
vast and the analysis often becomes a bottleneck in the processing pipeline
\supercite{rahmanianEnablingModularAutonomous2022}. Powder X-ray diffraction
(XRD) is an important measurement technique used to obtain structural
information from polycrystalline samples\supercite{harrisContemporaryAdvancesUse2001}. 
The diffractograms are an
information-dense fingerprint of the structure of the material. However,
analyzing these diffractograms is not an easy task\supercite{holderTutorialPowderXray2019}. 
Full structure solutions and
Rietveld refinement take time and require expert knowledge, both about the
analysis technique and the materials class at hand. This is not feasible in high-throughput experiments on a larger
scale. Therefore, the question arises whether it is possible to automatically 
analyze powder diffractograms with machine learning models trained on large
amounts of data, making it possible to run inference almost instantaneously.

During the last few years, there have been several studies tackling this
objective by applying machine learning models to various tasks concerning the
analysis of powder diffractograms such as phase
classification\supercite{leeDeeplearningTechniquePhase2020,maffettoneCrystallographyCompanionAgent2021,schuetzkeEnhancingDeeplearningTraining2021,szymanskiProbabilisticDeepLearning2021,wangRapidIdentificationXray2020},
phase fraction determination\supercite{leeDatadrivenXRDAnalysis2021}, space
group
classification\supercite{parkClassificationCrystalStructure2017,oviedoFastInterpretableClassification2019,zalogaCrystalSymmetryClassification2020,vecseiNeuralNetworkBased2019,suzukiSymmetryPredictionKnowledge2020,chakrabortyDeepCrystalStructure2022},
machine-learning-guided Rietveld
refinement\supercite{ozakiAutomatedCrystalStructure2020,fengMethodArtificialIntelligence2019},
extraction of lattice
parameters\supercite{dongDeepConvolutionalNeural2021,chitturiAutomatedPredictionLattice2021,chakrabortyDeepCrystalStructure2022,
habershonPowderDiffractionIndexing2004} and crystallite
sizes\supercite{dongDeepConvolutionalNeural2021,chakrabortyDeepCrystalStructure2022},
and also novelty detection based on unsupervised techniques\supercite{bankoDeepLearningVisualization2021}.
Since an abundant source of experimental diffractograms is hard to come by, most
applications train their models on simulated diffractograms from the Inorganic
Crystal Structure Database (ICSD)\supercite{bergerhoff1987}, which contains a total of 272\,260 structures (October 2022).

\citeauthor{leeDeeplearningTechniquePhase2020} used a deep convolutional neural
network (CNN) trained on a large dataset of multiphase compositions from the
quaternary \ce{Sr-Li-Al-O} pool to classify present phases in the
diffractogram\supercite{leeDeeplearningTechniquePhase2020}. In a follow-up
study, they further showed good results for phase fraction inference in the
quaternary \ce{Li-La-Zr-O} pool\supercite{leeDatadrivenXRDAnalysis2021}.
\citeauthor{schuetzkeEnhancingDeeplearningTraining2021} performed phase
classification on iron ores and cement compounds and used data augmentation with
respect to lattice parameters, crystallite sizes, and preferred orientation
\supercite{schuetzkeEnhancingDeeplearningTraining2021}. They showed that
especially the lattice parameter variations enhance the classification accuracy
significantly.

Instead of the analysis of phase composition,
\citeauthor{dongDeepConvolutionalNeural2021} performed regression of scale
factors, lattice parameters, and crystallite sizes in a five-phase catalytic
materials system\supercite{dongDeepConvolutionalNeural2021}. In contrast to
supervised tasks, \citeauthor{bankoDeepLearningVisualization2021} used a
variational autoencoder to visualize variations in space group, preferred
orientation, crystallite size, and peak
shifts\supercite{bankoDeepLearningVisualization2021}.
\citeauthor{parkClassificationCrystalStructure2017} used a deep CNN to classify
space groups of single-phase diffractograms, reaching a test accuracy of 81.14\%
on simulated diffractograms.\supercite{parkClassificationCrystalStructure2017} 
However, as we will show later in this paper, this accuracy is highly overestimated and drops to 56.1\% when test splits are designed in a way to reduce data leakage in non-IID datasets such as the ICSD.
\citeauthor{vecseiNeuralNetworkBased2019}\supercite{vecseiNeuralNetworkBased2019} developed
a similar approach and applied their classifier to experimental diffractograms
from the RRUFF mineral database\supercite{lafuentePowerDatabasesRRUFF2015},
reaching an experimental test accuracy of 54\%.

While the ICSD contains a large number of structures spanning many different
classes of materials, it still falls short in size, distribution, and generality
compared to the datasets used to train very large state-of-the-art models of
other fields such as computer vision.
Furthermore, the ICSD
database is highly imbalanced with respect to space groups, as can be seen
in the histogram in Figure~\ref{fig:histogram_dist_spgs}. This makes the classification of
space groups more difficult, as shown and discussed by
\citeauthor{zalogaCrystalSymmetryClassification2020}
\supercite{zalogaCrystalSymmetryClassification2020}. The ICSD also contains a
limited number of different structure types that may not adequately represent
the crystal structures analyzed in future experiments.

To overcome these shortcomings, we propose to train machine learning models on
diffractograms simulated from synthetic crystal structures randomly generated
based on the symmetry operations of the space groups. This makes it possible to
train on structures with new structure types not present in the ICSD. We used
the crystals from the ICSD only to determine vague statistics guiding the random
generation and for calculating the test accuracy. Our approach goes one step
further than classical data augmentation by fully detaching itself from the
individual entries in the ICSD database.
\replaced{The generated synthetic crystals form a training dataset that includes stable ICSD crystal structures, unstable crystal structures,
but also stable structures that are not yet present in the ICSD. By training a model on the full dataset, we can also expect improvements on the unknown stable crystal structures.}{It is our hypothesis that when trying
to solve more general tasks concerning the extraction of structural details from
powder diffractograms, the crystals from which the training dataset is simulated
do not need to be stable or chemically viable.} \replaced{Furthermore, we}{We} propose viewing the problem
\deleted{simply} as a mathematical task of getting back some of the real-space information
leading to given powder X-ray diffractograms. \added{Therefore, even the unstable structures included in our generated dataset will help to learn to classify the stable structures.}

Here, we applied this approach to the classification of the crystal symmetry,
namely the space group. The space group is usually one of the first structural
pieces of information needed after synthesizing a new material. This task is
well-suited to showcase the strengths of using a synthetic dataset and to
benchmark it. We further show the results of using our methodology to infer space group labels of an experimental dataset.

We embedded our synthetic generation algorithm in a framework with
distributed computing capabilities to generate and simulate diffractograms on
multiple nodes in parallel using the \emph{Python} library \emph{Ray}
\supercite{moritzRayDistributedFramework2018a}. In contrast to the traditional
approach of generating a simulated dataset before training, we used this
distributed computing architecture to build an infinite stream of \replaced{synthetically}{randomly}
generated and simulated diffractograms to perform batch-wise online learning.
This increases the generalization performance, eliminates the problem of
overfitting, and allows very large models to be trained.

\section{Methods}
\subsection{Generating synthetic crystals} \label{sec:generate_crystals} 

\begin{figure}[!htb] 
    \centering
    \includegraphics{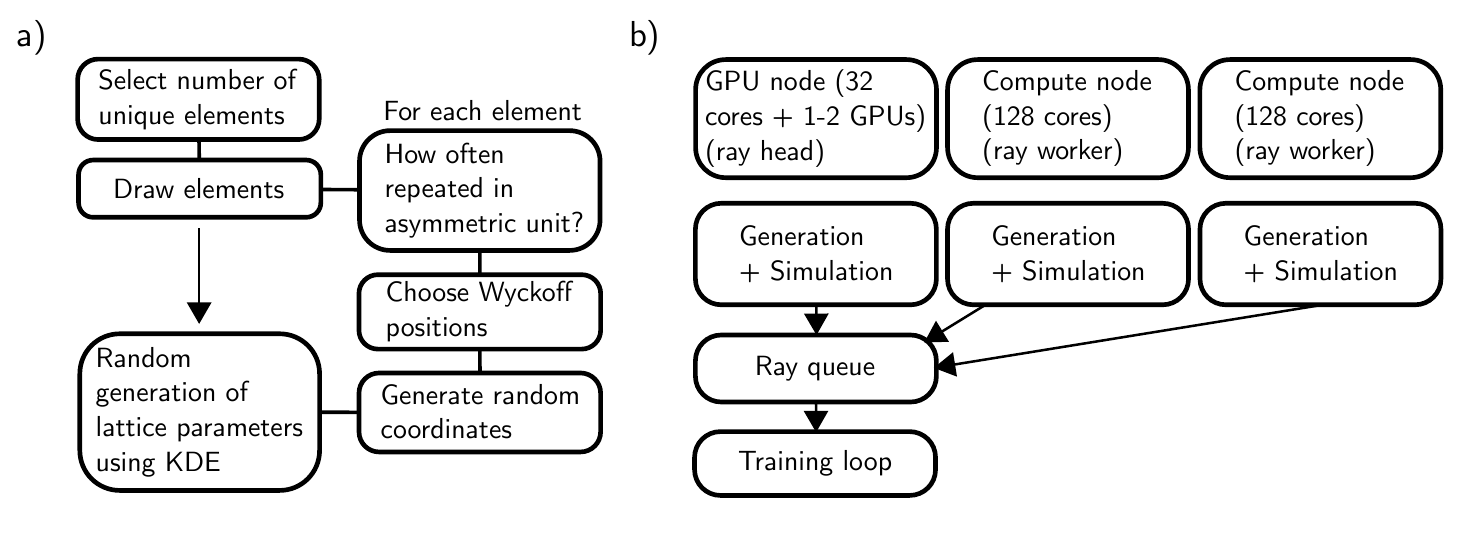}
    \caption{a) Flowchart of how the generation algorithm produces synthetic
    crystals. Atoms are independently placed on the Wyckoff positions and random
    coordinates are drawn. b) Overview of the distributed computing system
    implemented using the \emph{Python} library
    \emph{Ray}\supercite{moritzRayDistributedFramework2018a}. Two compute nodes
    (that generate and simulate diffractograms) are connected to the \emph{Ray}
    head node using a \emph{Ray queue} object.}
    \label{fig:generation_and_distributed}
    \end{figure}

    To generate synthetic crystals, we randomly place atoms on the Wyckoff
    positions of a given space group \added{following the Wyckoff occupation probabilities extracted from the ICSD} and
    then apply the respective symmetry operations. The algorithm is
    explained in the following (see also Figure~\ref{fig:generation_and_distributed}a for a flow diagram of the algorithm).
    We only explain the most important steps, details can be found in Section~S1 of the SI.

    \begin{enumerate}
        \itemsep0em
        \item \replaced{Sampling of a space group from the space group distribution of the ICSD.}{Random selection of a space group. We follow the space group distribution of the ICSD to allow comparison with previous work.}
        \item Sample unique elements of the crystal and their number of repetitions in the asymmetric unit.
        \item Place atoms onto the Wyckoff positions and draw uniform coordinates for each.
        \item Draw lattice parameters from a \replaced{kernel density estimate}{KDE} based on the ICSD.
        \item Apply space group symmetry operations.
    \end{enumerate}

    Parts of this algorithm were inspired by the generation algorithm of the
    \emph{Python} library \emph{PyXtal}
    \supercite{fredericksPyXtalPythonLibrary2021}. 
    We only keep generated crystals for training if the
    conventional unit cell volume is below \SI{7000}{\angstrom\cubed} and if
    there are less than 100 atoms in the asymmetric unit. 
    We did not employ any form of
    distance checks on the coordinates, as we found this to have no meaningful
    impact on space group classification accuracy. We only prevented the
    algorithm from placing more than one atom onto a Wyckoff position that does not
    have a degree of freedom. We also did not use partial occupancies. We chose
    this algorithm for its simplicity and its capability to reproduce many
    important characteristics of ICSD crystals adequately (see Section~
    \ref{sec:synth_distribution}).

    \begin{figure}[h]
    \centering
    \includegraphics{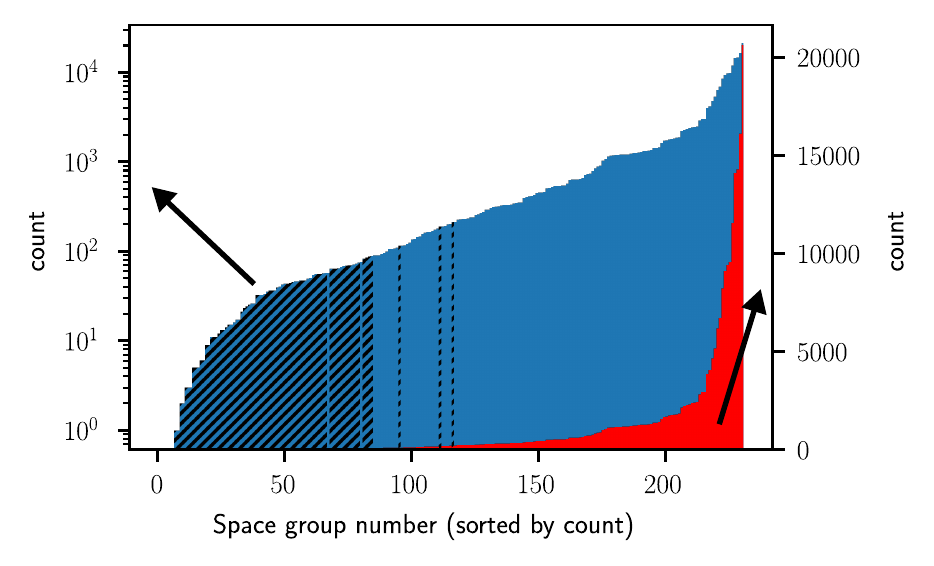}
    \caption{Distribution (logarithmic scale in blue, linear scale in red)
    of space groups in the ICSD. Space groups are sorted by count\added{ (see Figure S13 in the SI for the distribution without sorting by count)}. The population of the space groups varies by
    multiple orders of magnitude, showing that the ICSD is a highly imbalanced
    dataset regarding space groups. The space groups excluded due to insufficient
    statistics are visualized with black stripes. \replaced{The histogram displays the distribution of the full ICSD, while the exclusion of space groups that do not contain enough samples is based on the statistics dataset (which does not include the test dataset, see Section~\ref{sec:dataset_split}) that we used to guide the random crystal generation. Therefore, the excluded space groups are not exactly the first 85 counted from the left.}{Since the exclusion is based on 
    the statistics dataset and this histogram shows the full ICSD,
    the excluded space groups are not exactly the first 85 counted from the left.}}
    \label{fig:histogram_dist_spgs}
    \end{figure}
   
    For some space groups, there are not enough crystals in the ICSD to form a
    representative \replaced{kernel density estimate}{KDE} for the volume or to calculate suitable occupation
    probabilities for individual Wyckoff positions. Therefore, we chose to only
    perform the classification on space groups with 50 or more crystals
    available in the statistics dataset we used to extract the probabilities
    (see Section~\ref{sec:dataset_split}), resulting in the exclusion of 85
    space groups (see Figure~\ref{fig:histogram_dist_spgs}). A classifier trained directly on ICSD data of all space groups will likely not be able to properly identify these space
    groups containing very few samples.

    If a similar performance for all space groups is desired, a uniform
    distribution of space groups in the training dataset is needed. This is
    trivially possible with our synthetic approach, in contrast to training
    directly on the ICSD, where weighting, over-, or undersampling methods are
    needed \supercite{sunClassificationImbalancedData2009}. To allow a direct
    and fair comparison between our approach and the original approach of
    training directly on ICSD entries, we still followed the same distribution
    of space groups of the ICSD in our synthetic training dataset. This
    eliminates the problem that the effective number of total space groups is
    smaller when training on a highly imbalanced dataset, making it easier to
    reach high accuracies.

    \added{Our choice of not sampling the space groups uniformly and using
    general statistics extracted from the ICSD to guide the crystal generation
    algorithm further builds upon the hypothesis that future crystals will
    roughly follow the more general statistics already present in the ICSD. With
    the chosen crystal generation algorithm we tried to find a middle ground
    between being much more general than using the ICSD crystals directly and
    not being too general such that it is very hard to extract structural
    information at all.}

    \subsection{Simulating diffractograms} \label{sec:simulation} 

    To simulate powder X-ray diffractograms based on the generated crystals, we used
    the implementation found in the \emph{Python} library \emph{Pymatgen}
    \supercite{ongPythonMaterialsGenomics2013}. We optimized the simulation code
    using the LLVM just-in-time compiler \emph{Numba}
    \supercite{lamNumbaLLVMbasedPython2015}. This increases the performance of
    the main loop over the reciprocal lattice vectors of the crystal
    significantly and makes the continuous simulation while training (discussed
    in the next section) possible.

    We used the wavelength \SI{1.5406}{\angstrom} (\replaced{Cu K$\alpha_1$}{Cu K$_\alpha$} line) to
    simulate all diffractograms. The obtained peaks were further broadened with
    a Gaussian peak profile to form the full diffractogram. To obtain the peak
    widths, we used the Scherrer
    equation\supercite{gilmoreInternationalTablesCrystallography2019}
    \begin{equation}
        \beta=\frac{K \lambda}{L \cos \theta} \text{,} \label{eq:scherrer}
    \end{equation}
    where $ \beta $ is the line broadening at half maximum intensity (on the
    $2\theta$-scale), $K$ is a shape factor, $\lambda$ is the wavelength, and
    $L$ is the (average) thickness of crystallites. We drew crystallite sizes
    from the range $ \left[ 20, 100 \right] $ \si{\nano \metre} and used 
    $K=0.9$.

    Diffractograms were generated in the range $ 2 \theta \in \left[ 5, 90
    \right] \si{\degree} $ with step size \SI{0.01}{\degree}. After generating
    each diffractogram, it was rescaled to fit in the intensity range $ \left[
    0,1 \right]$. \added{In Figure S9 of the SI we show an exemplary diffractogram simulated 
    from the ICSD, Figure S10 shows an exemplary diffractogram simulated from a synthetic 
    crystal.}

    \subsection{Continuous generation of training data}\label{sec:online_learning} 
    
    Typically, machine learning models are trained with a fixed dataset predefined at the beginning of training. Sometimes, data augmentation
    is applied to further increase the effective size of this dataset. In
    contrast to that, we generated our dataset on-the-fly, parallel to model training. The
    main advantage of using this approach compared to a fixed-size dataset is
    the eliminated possibility to overfit to individual diffractograms since
    every diffractogram is only used once. Furthermore, not having to
    pre-simulate a dataset before training makes this approach more flexible
    when changing simulation parameters.

    We used a distributed architecture on multiple nodes using the \emph{Python}
    framework \emph{Ray} \supercite{moritzRayDistributedFramework2018a}, which enabled the training on 1-2 GPUs and simultaneous generation of training data on more than 200 CPU cores (see Figure~\ref{fig:generation_and_distributed}b and SI Section~S2.2). 
    Depending on
    the model size and corresponding training speed, this setup allows
    training with up to millions of unique diffractograms per hour.

    \subsection{Dataset split} \label{sec:dataset_split} 
    
    The ICSD database contains many structures that are very similar with
    slightly different lattice parameters and coordinates. For example, there
    are 25 entries for \ce{NaCl} (October 2022). Furthermore,
    there are 3898 entries that have the same structure type as \ce{NaCl} and
    thus also similar powder diffractograms. If some of them appear in the
    training dataset and some in the test dataset, the classification will be
    simplified to recognizing the structure type or structure. In that case, the
    test set accuracy will not represent the true generalization performance of the neural
    network. To quantify the true generalization performance, we split the
    dataset in such a way that the same structure type appears either only in
    the training or in the test dataset. We used the structure type definitions
    provided by the ICSD. The obtained accuracy on the test dataset reflects the
    accuracy of our network when being used on a novel sample with a structure
    type not yet present in the ICSD database.

    We want to emphasize that the used test split is very important for the task of
    space group classification and not a trivial choice. The ICSD contains many
    subtypes of structure types (for example, subtypes of perovskites), which we
    regarded as separate structure types in our split. Considering the subtypes
    as the same structure type may also be a viable option when performing the
    split. A combination of a split based on structure type and sum formula or
    similar approaches are also possible. \deleted{However, we think that our choice of
    splitting the dataset is able to effectively measure the generalization
    performance while still being relatively simple.}

    \added{Depending on the experimental setting, it further might make more sense in some cases to not do a structure type-based split. If the likelihood of finding structures similar to already-discovered structure types in the planned experiment is high, training should definitely include those structure types to evaluate the performance of the model. However, in a pure discovery setting, new structure types can appear. To evaluate the expected model performance in this scenario and thus quantify the true generalization error to unseen data, we chose the most strict structure type-based split.}

    We divided the ICSD (database version 2021, June 15) in a 70:30 split. For
    our synthetic crystal approach, the 70\% part (\replaced{which we
    call}{called} statistics or training dataset) was only used to create the
    \replaced{kernel density estimates}{KDEs} and to calculate the Wyckoff
    occupation probabilities needed for the generation algorithm. Since we can
    judge the performance of the synthetic generation algorithm by comparing the
    training accuracy (on synthetic crystals) with the accuracy tested on
    diffractograms simulated directly from the statistics dataset, an additional
    validation dataset was not needed. For comparison with the original approach
    of directly training on ICSD
    crystals\supercite{parkClassificationCrystalStructure2017}, we simulated
    crystals directly from the statistics dataset and trained on them.

    Analogous to the synthetic generation, we only used crystals with a
    conventional unit cell volume below \SI{7000}{\angstrom\cubed} and with less
    than 100 atoms in the asymmetric unit for the statistics and test dataset.
    This covers $ \approx 94\% $ of the ICSD crystals.

    \subsection{Models and \added{computational} experiments} \label{sec:models_and_experiments}

    \subsubsection*{Models}
    We will briefly introduce the models we used for the classification
    of space groups. A more detailed description can be found in the SI
    Section~S2.1.

    As a baseline, we first used the CNN models introduced by
    \citeauthor{parkClassificationCrystalStructure2017}\supercite{parkClassificationCrystalStructure2017}.
    They used three models, one for the classification of crystal systems
    (``parkCNN small''), one for extinction groups (``parkCNN medium''), and one
    for space groups (``parkCNN big''). All models have three convolution layers
    with two hidden fully connected layers and one output layer. The
    three models differ in the number of neurons in the hidden fully connected
    layers, increasing the number of model parameters with the number of target
    labels. Here, we only used the models ``parkCNN medium'' and ``parkCNN big''
    and applied both to the classification of space groups. When using ICSD
    crystals to train the ``parkCNN'' models, dropout was used, while the
    training of the ``parkCNN'' models on synthetic crystals did not use
    dropout.

    Since the approach of using an infinite stream of generated training data
    eliminates the problem of overfitting, we further used deeper models with a
    higher number of model parameters. For this, we used the deep convolutional
    neural networks ResNet-10, ResNet-50, and ResNet-101, which were introduced
    by
    \citeauthor{heDeepResidualLearning2016}\supercite{heDeepResidualLearning2016}
    in 2015.

    Details of the machine learning setup can be found in the SI Section~2.2. Overall, our setup allowed us the training of models over up to 2000 epochs with more than $100\,000$ unique, newly generated crystals and corresponding diffractograms in each epoch \added{(see the upper x-axis of Figure \ref{fig:training_curve})}.

    \subsubsection*{\replaced{Computational experiments}{Experiments}}

    We performed two sets of experiments to evaluate our new dataset split as well as our synthetic crystal generation approach and compare it to state-of-the-art models in literature: Firstly, we trained and tested models on ICSD crystals only, and secondly, we trained on synthetic crystals and tested on ICSD crystals.

    In particular, we first performed an experiment with the ``parkCNN medium''
    model trained directly on the diffractograms simulated from the ICSD
    statistics dataset with a fully random train-test split (similar to
    \supercite{parkClassificationCrystalStructure2017}), instead of splitting by
    the structure type of the crystals. This experiment makes a comparison of
    the two different methods of train-test split possible. We then trained the
    ``parkCNN big'' model using the structure type-based split, again directly
    on ICSD diffractograms. We further repeated the same experiment using the
    smaller model ``parkCNN medium'' to resolve potential overfitting to the
    ICSD diffractograms.

    For the experiments performed on our continuously generated
    dataset based on synthetic crystals, we used the structure type-
    based split. As discussed in Section~\ref{sec:dataset_split}, the training / 
    statistics dataset was only used to extract more general statistics, such as the 
    element distribution.
    First, we trained the ``parkCNN big'' model. For each
    batch, we generated $ 435 $ random crystals and simulated two diffractograms
    with different crystallite sizes for each of them. This resulted in the
    batch size of $ 870 $. 
    Since our synthetic crystal generation algorithm yields an infinite stream
    of unique diffractograms to train on, using much larger models than for the
    fixed ICSD dataset is possible without overfitting. We performed experiments
    for the ResNet-10, ResNet-50, and ResNet-101 models. Instead of generating
    two diffractograms with different uniformly sampled crystallite sizes for each generated crystal (as
    we did for the ``parkCNN big'' model), we now created only one diffractogram
    for each of the $ 145 $ crystals generated for one batch. This is due to 
    the slower training of the ResNet models, which means that reusing the same 
    diffractogram with different crystallite sizes is not necessary to generate
    training data fast enough.
    
    To obtain the
    highest-possible ICSD test accuracy, we further applied the square root
    function as a preprocessing step to the input diffractograms of the
    network when using the ResNet models. This was suggested by
    \citeauthor{zalogaCrystalSymmetryClassification2020}
    \supercite{zalogaCrystalSymmetryClassification2020} and in their case
    improved classification accuracy by approximately 2 percentage points. Some
    initial tests suggested that this approach also yields a higher accuracy in our
    case, so we used this preprocessing step to train the ResNet
    models.

    While we focused mainly on the methodology of using synthetic crystals to
    extract structural information from powder diffractograms, we also show some
    initial steps toward applying our methods to experimental data. We used the
    publicly available RRUFF mineral database
    \supercite{lafuentePowerDatabasesRRUFF2015} which provides experimental
    measurements, including powder diffractograms \added{(see Figure S11 in the
    SI for an exemplary diffractogram from the RRUFF)}. In order to imitate
    experimental diffractograms, we added Gaussian additive and multiplicative
    noise \added{(similar to
    \supercite{vecseiNeuralNetworkBased2019,szymanskiProbabilisticDeepLearning2021})}
    and a background function based on samples from a Gaussian process to our
    simulated diffractograms. Furthermore, we added a small amount of an
    impurity phase to each diffractogram. Details about the experimental data
    generation protocol can be found in the SI Section~S4\added{, Figure S12
    shows an exemplary synthetic diffractogram with noise, background and an
    impurity phase.} Using the ResNet-50 model, we performed two experiments for
    experimental data, one with the mentioned impurity phase, and one without.

\section{Results and discussion}
    \subsection{Synthetic distribution} \label{sec:synth_distribution}

    \begin{figure}[!htb]
    \centering
    \includegraphics{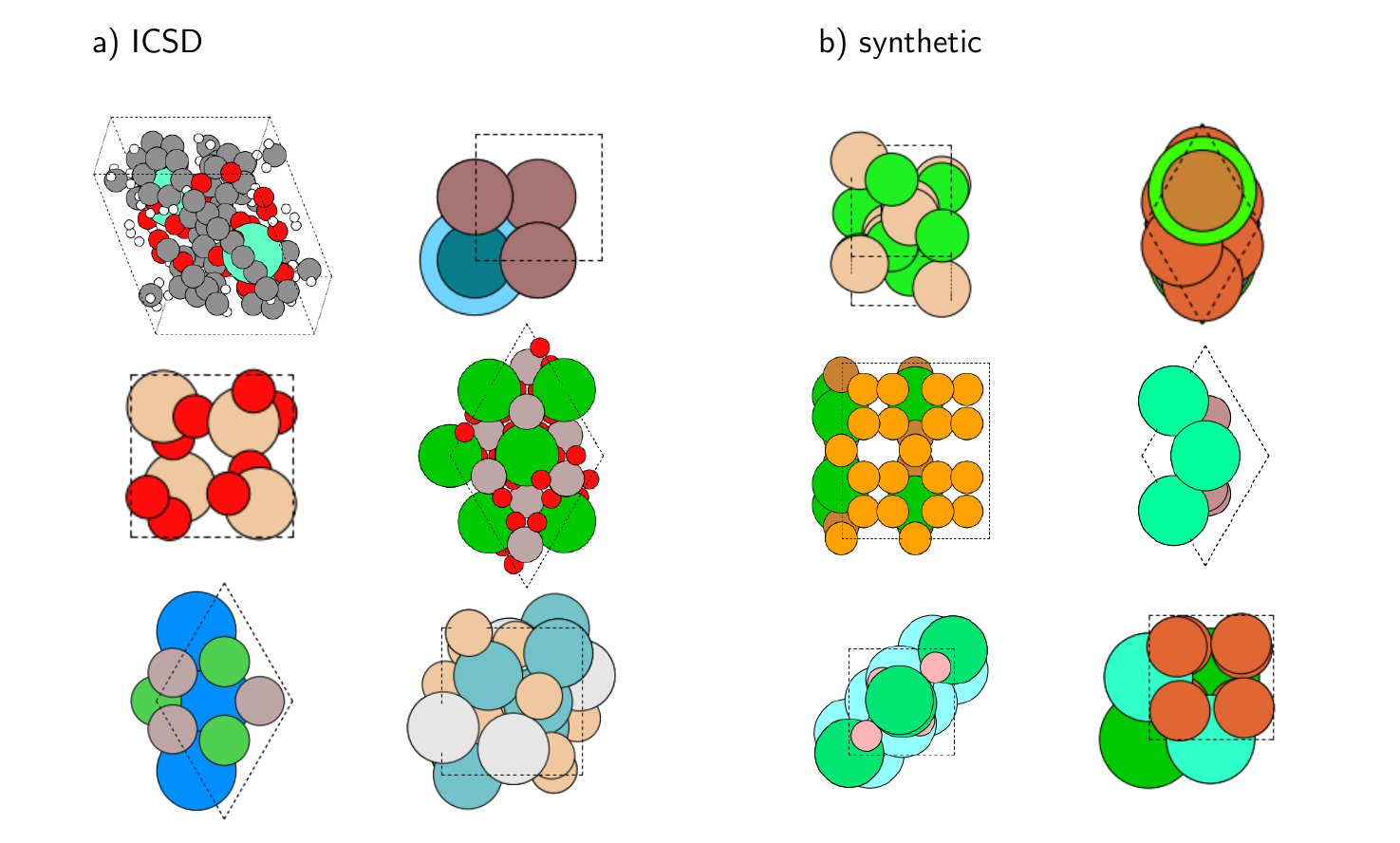}
    \caption{a) Some \replaced{randomly chosen and thus representative}{randomly picked} examples of ICSD crystals. b) Some \replaced{randomly chosen and thus representative}{randomly
    picked} examples of synthetically generated crystals. While coordination and
    distances are not chemically correct for the synthetic crystals, crystal
    symmetries are reproduced correctly.}
    \label{fig:example_structures}
    \end{figure}

    \begin{figure}[!htb]
    \centering
    \includegraphics{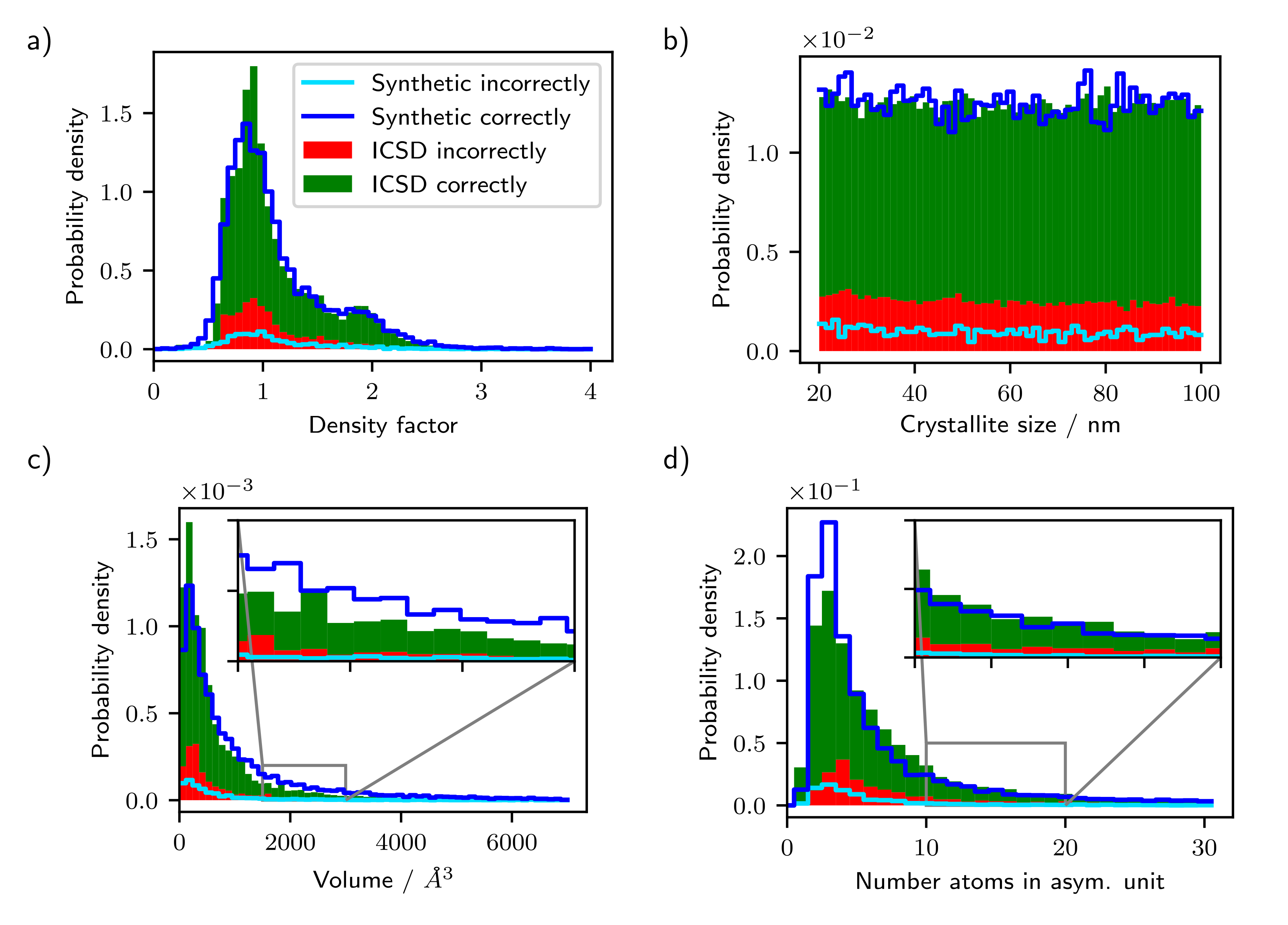} 
    \caption{Histograms comparing the distributions of descriptors of the
    synthetically generated crystals with the ICSD distribution in the test
    dataset. a) density factor $ \frac{V_\text{unit cell}}{\sum_i 4/3 \pi
    \left(\frac{r_{i\text{;cov}} + r_{i\text{;VdW}}}{2}\right)^3} =
    \frac{V_\text{unit cell}}{V_\text{atomic}}$, b) crystallite sizes, c) unit cell
    volume (conventional cell settings), d) number of atoms in the asymmetric
    unit. The probability density of the ICSD is visualized by a stacked bar
    histogram, where the green portion of the bar was correctly classified and
    the red portion was incorrectly classified. The probability density of the
    synthetic crystals is visualized by the dark blue line. The portion between the
    dark blue line and the light blue line was correctly classified, the portion below the
    light blue line was incorrectly classified. The reported classification
    performance is based on the ResNet-101 model trained on diffractograms from
    synthetic crystals.}
    \label{fig:histograms}
    \end{figure}

    We first present an analysis of the generated synthetic crystals.
    Figure~\ref{fig:example_structures} shows some \replaced{randomly chosen and thus representative examples of}{randomly selected} ICSD and
    synthetic crystal structures side-by-side. Visually, the crystals appear very
    similar. However, no physical or chemical considerations regarding stability, clashing atoms, and element combinations are taken into account in the generation of synthetic crystals. As discussed earlier, our goal is to demonstrate that this is
    not problematic when using these crystals for the extraction of structural
    information from powder diffractograms. On the contrary, we expect the synthetic crystals to be a better basis for generalization to fundamentally new crystal structures than existing finite databases.

    To compare the distribution of ICSD crystals with the synthetic
    distribution, we evaluated structural descriptors, i.e. density factors, crystallite sizes, unit cell volumes, and numbers of atoms in the asymmetric unit, and compare their histograms
    in Figure~\ref{fig:histograms}.
    One can see that the overall distributions of
    the synthetic and ICSD crystals are very similar for all four descriptors.
    This shows that our chosen generation algorithm reproduces crystals that are
    similar to ICSD crystals in terms of these more general descriptors.

    \subsection{Classification results} \label{sec:classification_results} 
    
    The main results of our experiments (see Section~\ref{sec:models_and_experiments}) to
    classify the space group of powder diffractograms can be
    found in Table~\ref{tab:results}. \added{In SI Table S2, we further provide the training time and total number of unique diffractograms for each computational experiment.}
    The goal of our experiments is to systematically analyse and quantify the changes in classification accuracy introduced by our two main contributions: A more challenging dataset split, and training on continuously generated synthetic data.

    We started by repeating previously reported
    experiments\supercite{parkClassificationCrystalStructure2017} trained
    directly using ICSD crystals with a random train-test split instead of the
    split based on structure types. This model achieved a very high test
    accuracy of 83.2\%. We note that the previous publication that we compare
    our results to\supercite{parkClassificationCrystalStructure2017} removed
    data from the training dataset, ``[...] heavily duplicated data
    [...]''\supercite{parkClassificationCrystalStructure2017}, but did not
    specify the exact criterions used. In contrast, we did not exclude any
    duplicates in this experiment based on a random train-test split.
    Furthermore, as discussed in Section~\ref{sec:generate_crystals}, we
    excluded crystals with a very high unit cell volume and a very high number
    of atoms in the asymmetric unit. This is likely the reason for the slightly
    higher classification accuracy that we observed, compared to the originally
    reported 81.1\%. 
    
    When splitting randomly, the model merely needs to recognize structures or
    structure types and assign the correct space group. This task is much easier
    than actually extracting the space group using more general patterns. When
    going from random splits to structure type-based splits (see
    Section~\ref{sec:dataset_split}), it becomes obvious that both the ``parkCNN
    big'' as well as the ``parkCNN medium'' models overfit the training data and
    do not generalize well to unseen structure types in the test set \added{(see
    Table \ref{tab:results})}. \added{The ``parkCNN medium'' model, which
    achieved 83.2\% on a random split, now only yields 55.9\% with the structure
    type-based split.}

    \replaced{Training the models by
    \citeauthor{parkClassificationCrystalStructure2017}}{Training the same
    model}, in particular the ``parkCNN big'' model, on synthetic crystals leads
    to a 1.6 percentage points higher test accuracy than the ``parkCNN big''
    model trained on ICSD diffractograms. At the same time, the training
    accuracy drops from \added{the} 87.2\% \added{when we trained the model
    directly on the ICSD} to 74.2\% \added{on the synthetic distribution}
    indicating that the model is now limited more by missing capacity rather
    than by overfitting, which is why we explored larger models, which will be
    discussed later. The gap between training and test accuracy is $31.1$
    percentage points when training on ICSD data, while for training using
    synthetic crystals, the gap is only $16.5$ percentage points. We note that
    this gap between training using synthetic crystals and testing using ICSD
    crystals cannot stem from overfitting, since no diffractograms are repeated
    for the synthetic training. The difference rather stems from the differences
    between the synthetic distribution and the ICSD distribution of crystals.

    \begin{table}[!htb]
        \begin{center}
        \caption{Results of training on diffractograms simulated from ICSD crystals (random splits as well as structure type-based splits) compared to when training on diffractograms from synthetic
        crystals. Test accuracy
        in all cases refers to the accuracy when testing on the ICSD test
        dataset. The training accuracies are averaged over the last 10 epochs of
        the respective run.
        Experiments trained directly on ICSD data
        overfitted to the training data. Training longer would have further
        increased the training accuracy, while not increasing the test
        accuracy.}
        \vspace*{2mm}
        
        \begin{tabular}{ccccccc} 
            \toprule
            Split & \begin{tabular}{@{}c@{}}Training \\ dataset\end{tabular} & \begin{tabular}{@{}c@{}}Testing \\ dataset\end{tabular} & Model & \begin{tabular}{@{}c@{}}Number of \\ parameters\end{tabular}& \begin{tabular}{@{}c@{}}Training acc. \\ / \%\end{tabular} & \begin{tabular}{@{}c@{}}Test acc. \\ / \%\end{tabular} \\
            \midrule
            Random & ICSD & ICSD & parkCNN medium & 4\,246\,797 & 88.4 & 83.2 \\
            \midrule
            Structure& \multirow{2}{*}{ICSD} & \multirow{2}{*}{ICSD}& parkCNN big & 4\,959\,585 & 87.2 & 56.1 \\
            type&&& parkCNN medium & 4\,246\,797 & 90.9 & 55.9 \\
            \midrule
            & \multirow{4}{*}{synthetic} & \multirow{4}{*}{ICSD} &parkCNN big & 4\,959\,585 & 74.2 & 57.7 \\
            Structure&&& ResNet-10 & 9\,395\,025 & 87.2 & 73.4 \\
            type$^{1}$ &&& ResNet-50 & 41\,362\,385 & 91.8 & 79.3 \\
            &&& ResNet-101 & 60\,354\,513 & 92.2 & \textbf{79.9} \\
            \bottomrule
        \end{tabular}\label{tab:results}
        \end{center} 
        \footnotesize{$^{1}$Here, the split type refers to the statistics and the test dataset, rather than the training and the test dataset.}
    \end{table}

    While the ``parkCNN big'' model trained on synthetic crystals outperforms
    the approach of training directly on ICSD crystals by only 1.6 percentage
    points, the advantage of training on an infinite stream of synthetic data increases when using models with more parameters and thus higher capacity. In
    contrast to training directly on a finite set of ICSD crystals, it is possible to train
    very large models using the infinite synthetic data stream without the
    potential of overfitting. As found in the last lines of Table~\ref{tab:results}, ResNet-10, ResNet-50, and ResNet-101 based models achieve ICSD test accuracies of 73.4\%,
    79.3\%, and 79.9\%. This is a significant increase from the 57.7\% achieved
    by the ``parkCNN big'' model. Figure~S4 in the SI further shows the top-$k$ accuracy over $k$ for the ResNet-101 model. With increasing $k$ the accuracy exceeds 95\% at $k=5$.
    This means that our model can not only determine the correct space group with a high probability but can also generate an almost complete list of possible space group candidates.
    
\begin{figure}[!htbp] 
    \centering
    \includegraphics{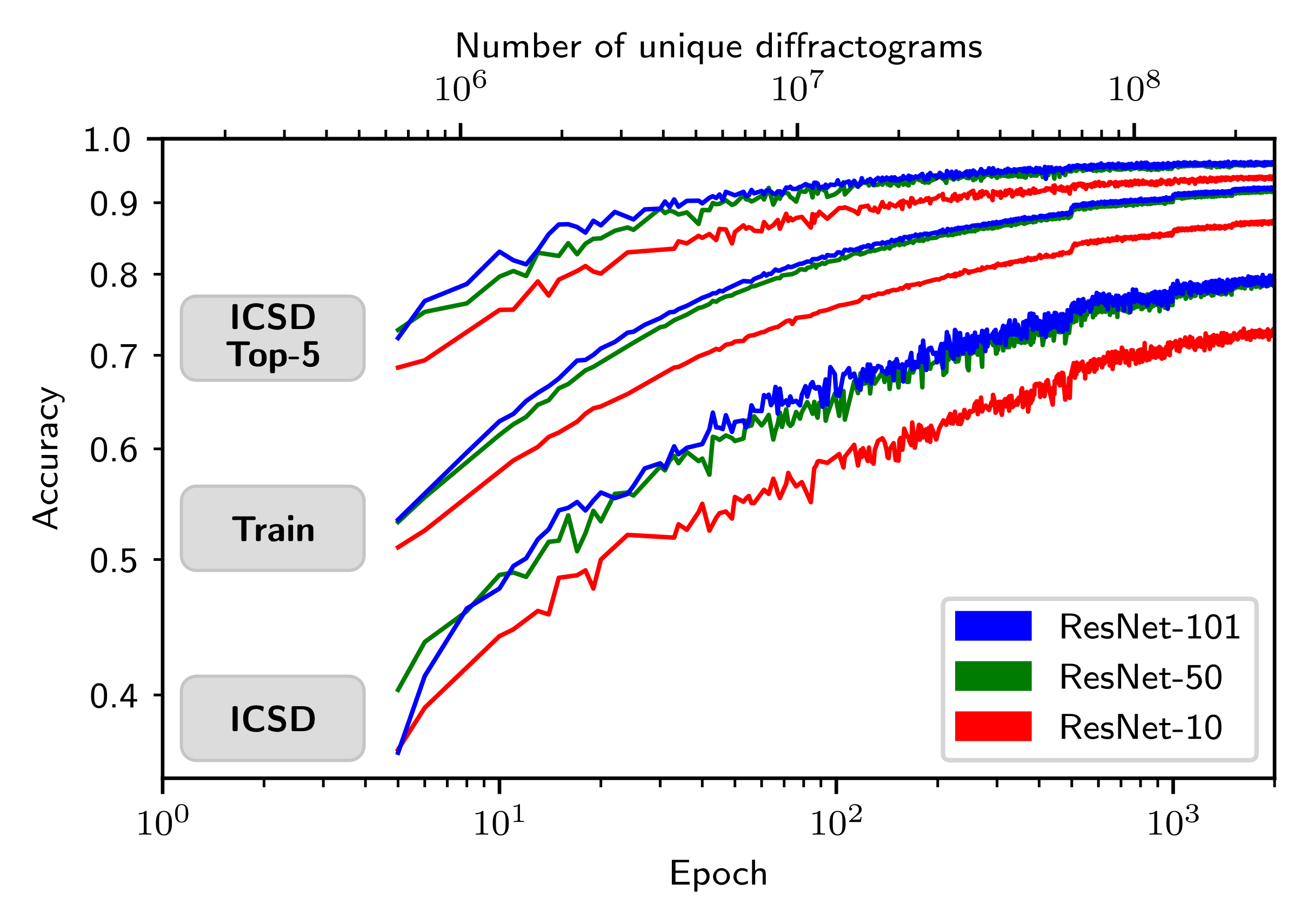}
    \caption{Test accuracy (ICSD), training accuracy (synthetic crystals), and
    test top-5 accuracy (ICSD) as a function of epochs\added{ (bottom axis). Since each additional epoch contains newly generated unique diffractograms, we further show the accuracies as a function of the total number of unique synthetic diffractograms (top axis)}. We show all three
    metrics for the models ResNet-101, ResNet-50, and ResNet-10. To better show
    the scaling behavior, both axes use logarithmic scaling. \added{Figure S6
    shows the same plot but without logarithmic scaling.} To better see the
    exponential behaviour, see Figure~S5 in the SI. }
    \label{fig:training_curve} 
    \end{figure}

    Figure~\ref{fig:training_curve} shows the ICSD test accuracy, the training
    accuracy (on synthetic data), and the ICSD top-5 test accuracy for all three ResNet variants
    as a function of epochs trained. For all three metrics, the difference
    between ResNet-50 and ResNet-101 is comparably small, while the step from
    ResNet-10 to ResNet-50 is substantial (5.9 percentage points in ICSD test
    accuracy, see Table~\ref{tab:results}). This shows that going beyond the
    model size of the ResNet-101 will likely not yield a big improvement in
    accuracy. In contrast to the 79.9\% accuracy reached in the top-1 ICSD test accuracy, the top-5 ICSD test accuracy of the ResNet-101 model reaches
    96\%. However, for all three ResNet variants, a gap between training using synthetic
    crystals and testing using the ICSD remains (12.3 percentage points for
    ResNet-101). As also shown in Figure~S5 in the SI, the accuracy convergence can be approximately described by a power law, indicating that exponentially more training time will substantially reduce classification errors and thus potentially lead to top-1 accuracies of 90\% and above, at the cost of a 100-fold increase in training times\replaced{. Considering the current training times provided in Table S2 of the SI, this is currently infeasible or only possible with tremendous hardware resources.}{, which is currently infeasible.}

    The histograms in Figure~\ref{fig:histograms} show, next to the overall
    distribution, also the fraction of diffractograms classified wrongly for
    testing on the ICSD (red bar) and on the synthetic data (below the light blue line)
    for the ResNet-101 model. First, one can see that throughout almost all
    regions of the distributions, the accuracy on the synthetic data is slightly
    higher than that on the ICSD. This is related to the aforementioned gap of 12.3
    percentage points between train and test accuracy and can be attributed to
    differences between the synthetic and ICSD distribution of crystals. This
    will be discussed in detail in the next section. It is surprising to see that the
    dependence on crystallite sizes is rather weak, as smaller crystallite sizes
    result in broader peaks (see Scherrer equation, Eq.~\ref{eq:scherrer}), potentially making the
    classification harder due to more peak overlaps.

    In summary, the maximum ICSD test accuracy of 79.9\% that we achieved using
    the ResNet-101 model almost reaches the previously
    reported\supercite{parkClassificationCrystalStructure2017} 81.14\% for the
    space group classification. However, our accuracy is based on a train-test
    split based on structure types, in contrast to a random split. This creates
    a much harder but also realistic task to solve since the model needs to
    generalize to other structure types without merely recognizing
    diffractograms or structure types that it has already seen during training.
    This becomes especially apparent from our experiment directly trained on
    diffractograms from ICSD crystals with the split based on structure types,
    which reached only 56.1\% instead of the previously
    reported\supercite{parkClassificationCrystalStructure2017} 81.14\%.

    \subsubsection*{Experimental results}
    To go beyond simulated diffractograms, we trained ResNet-50 models on
    calculated diffractograms with background, noise, and impurities and applied
    the trained models to the RRUFF mineral database. Our results (see Figure~S3
    in the SI) show that it is essential to include impurity phases in the
    training data. By doing so, we obtain a top-1 accuracy of 25.2\% and a
    top-10 accuracy of over 60\%. This is of high practical relevance since
    having a short list of potential space groups is often sufficient as a first
    step to further refinement and analysis.

    \citeauthor{vecseiNeuralNetworkBased2019} performed similar experiments of
    space group classification on the same database. Using an ensemble of 10
    fully connected neural networks, they reached a classification accuracy of
    54\%\supercite{vecseiNeuralNetworkBased2019}. While our obtained accuracy is
    significantly lower, our approach is much more general: In contrast to our
    approach, the training dataset was based on simulated diffractograms of
    structures of the ICSD\supercite{vecseiNeuralNetworkBased2019}, which
    contains almost all RRUFF structures, leading to high similarities of
    training and \replaced{test}{testing} data. Therefore, the model needed to
    simply recognize the minerals, instead of directly inferring the space group
    using the symmetry elements - as our method needs to do.

    We want to emphasize that our efforts to apply the methodology to
    experimental data are only preliminary. We expect improved results with an
    improved data generation protocol since the procedure contains many
    parameters to be tuned. Ideally, one would use a generative machine learning
    approach to add the experimental effects (noise, background, impurities) to
    the pure diffractograms. We also want to point out that the noise level and
    quality of data in the RRUFF dataset are limited. Application of the
    presented methodology \replaced{to}{on} other experimental datasets is
    desirable. \deleted{We will address these points in future work, where we
    focus on improved ways of modeling experimental imperfections.}
    \added{As discussed above, for the
    classification of pure diffractograms we observed the ResNet-50 to have the
    best cost-benefit ratio, since the ResNet-101 yielded only slight
    improvements. For the more complicated problem of classifying diffractograms
    with experimental imperfections, bigger models and longer training times
    might be necessary.}
    
    \added{Next to improving the modeling of experimental imperfections and therefore the overall accuracy on experimental data, the practical application of deep neural networks for analyzing powder diffractograms yields further challenges that we want to discuss. Since experimental setups differ, e.g., concerning the used wavelength, a different $2\theta$ step size, or a different $2\theta$ range, a new neural network would need to be trained for each situation. Since our largest model requires a significant computational investment, this might not be feasible in all situations. Arguably, though, for large high-throughput experiments, the 11-day training of a ResNet-50 should not be unreasonable, especially if it can speed up the data analysis significantly and allow in-loop adaptive experimentation. For smaller setups, where this is not feasible, other solutions must be found. First, one can use a form of transfer learning from a pre-trained model to fine-tune to the desired experimental setup. This, however, would only work for a change in wavelength, since a change in step size or $2\theta$ range would change the input dimensions of the network. However, to handle a change in the $2\theta$ range, it might be possible to include a form of zero-masking in the synthetic training data, such that different input ranges (with zeros where no measurement was made) can be used, which would lead to a more flexible model, not requiring new training data when applied to a new $2\theta$ range. For a change in the step size, a cubic spline interpolation might be helpful. We plan to address these challenges in future work.}

    \deleted{As discussed above, for the classification of pure diffractograms we
    observed the ResNet-50 to have the best cost-benefit ratio, since the
    ResNet-101 yielded only slight improvements. For the more complicated
    problem of classifying diffractograms with experimental imperfections,
    bigger models and longer training times might be necessary.}

    \added{Furthermore, analysis of the loss value or gradient norm associated with particular samples, i.e. crystal structures, during training on synthesis crystals or during transfer learning from synthetic to experimental data can help to better understand the relevance and informativeness of given samples for the model. This can help in generating more relevant synthetic data based on experimental crystal structures that are underrepresented in the synthetic data distribution.}

    \subsection{Differences between synthetic crystals and ICSD crystals}
    We showed that training directly on crystals from the ICSD yields a gap
    between the training and test accuracy due to overfitting. The training on
    the synthetic dataset also shows a gap between the training and test
    accuracy (see Table~\ref{tab:results}), but it is smaller than when training
    directly on ICSD crystals. Furthermore, this gap is not due to overfitting,
    since overfitting to singular diffractograms is not possible when the model
    is trained using an infinite stream of generated synthetic crystals. The gap
    rather stems from systematic differences between the synthetic and ICSD
    distribution of crystals.

    To analyze those differences, we created three modifications of the ICSD
    test dataset \added{(see SI Section~S3 for details)}. In the first modification, the fractional coordinates of the
    atoms in the asymmetric unit of the crystals of the ICSD test dataset were
    randomly uniformly resampled (as in the synthetic crystal generation
    algorithm). In the second modification, the lattice parameters were
    randomized following the \replaced{kernel density estimate}{KDE} used in the synthetic generation algorithm. The
    third modification combines both previous modifications, i.e. both the
    coordinates and the lattice parameters were resampled. These three modified
    test datasets bring the ICSD test dataset closer to the distribution used
    for training and let us quantify which factors contribute to the gap between
    training \added{on synthetic crystals} and testing \added{on the ICSD}. 

    We evaluated the test accuracies on \replaced{the randomized}{these}
    datasets for the experiment using the ResNet-101 model trained using
    synthetic crystals. We found that randomizing the coordinates yields an
    increase in test accuracy of \replaced{$4.89$}{$5.96$} percentage points.
    Randomizing the lattice parameters results in an increase of
    \replaced{$0.79$}{$1.31$} percentage points. Randomizing both the
    coordinates and the lattice parameters leads to an increase of
    \replaced{$5.70$}{$6.32$} percentage points, explaining
    \replaced{almost}{approximately} half of the gap of 12.3 percentage points
    between synthetic training and ICSD test accuracy.

    \added{So far, we have randomized the lattice parameters and coordinates of the test dataset, such that they follow a distribution that is based on the statistics extracted from the statistics dataset. However, this does not take into account the different Wyckoff position occupation probabilities between the test and statistics datasets. For this, we repeated
    a similar analysis, for which we applied the
    randomizations to the statistics dataset rather than the test dataset.
    Without any modifications, testing on the statistics dataset instead of the test dataset yielded $3.89$ percentage points higher accuracy. This can be explained by slight differences in the overall statistics between the test and statistics datasets.
    Randomizing the coordinates yields a further increase of $4.72$
    percentage points, randomizing the lattice $1.16$ percentage points, and
    randomizing both the coordinates and the lattice parameters $6.68$ percentage points.
    In total, testing on the statistics dataset with randomized coordinates and
    lattice parameters yields a $10.57$ percentage points higher accuracy than on the
    unmodified test dataset. This almost completely explains the gap
    of $12.3$ percentage points between the training accuracy on synthetic crystals and the test accuracy on the ICSD. The
    remaining part is likely due to our algorithm that places atoms on Wyckoff
    positions 
    not reproducing the ICSD distribution exactly. 
    However, the
    remaining difference is remarkably small.}

    \deleted{We also tested the ResNet-101 model on diffractograms simulated
    based on the statistics dataset from which the statistics for our generation
    algorithm have been extracted. This returns a $4.33$ percentage points
    higher accuracy than on the test dataset. This is due to the different
    structure types present in the test and statistics dataset yielding
    different occupation probabilities (potentially also 0) for Wyckoff
    positions and overall slightly different statistics for the synthetic
    generation algorithm.}
    \deleted{To summarize, the contributions to the gap stem from differences in the
    distribution of lattice parameters and coordinates with $6.32$ percentage
    points and from the differences between the statistics dataset and test
    dataset with $4.33$ percentage points. Added together (assuming the effects of geometry and Wyckoff position occupation probabilities are unrelated and thus additive), the contributions
    almost completely explain the gap of $12.3$ percentage points between
    training and test accuracy. The remaining part is likely due to our
    algorithm that places atoms on Wyckoff positions not reproducing the ICSD
    distribution exactly. However, the remaining difference is remarkably small.}

    \begin{figure}[!htb]
    \centering
    \includegraphics{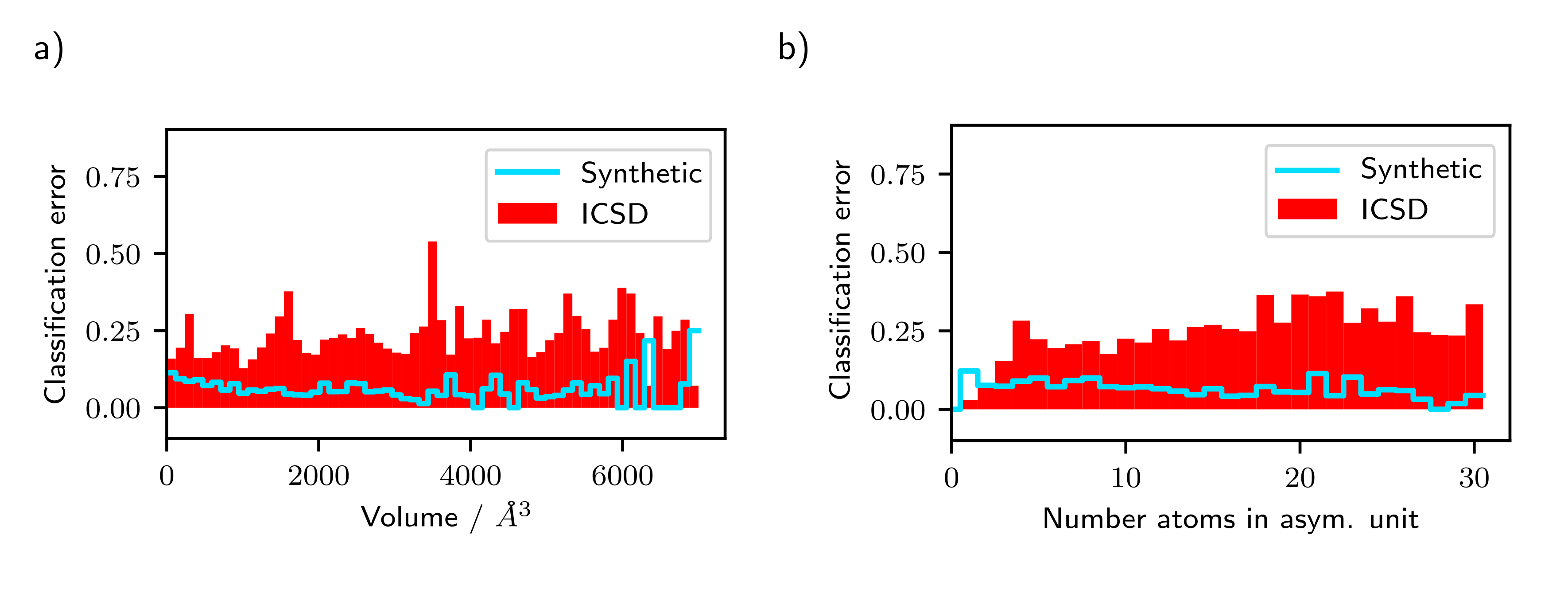}
    \caption{Classification \replaced{error}{performance} for each bin of a) the unit cell volume
    (conventional cell settings) and b) the number of atoms in the asymmetric
    unit. \deleted{The classification performance when testing on the ICSD is shown using
    a stacked bar plot, where the red bar indicates the classification error and
    the green bar indicates the correctly classified portion of each bin. When
    testing on the synthetic distribution, the classification error is given by
    the light blue line, the rest (between dark blue and light blue line) was correctly
    classified.} The reported classification performance is based on the
    ResNet-101 model trained on diffractograms from synthetic crystals. This
    visualization clearly shows the error rate within each bin, in contrast to
    Figure~\ref{fig:histograms}, which additionally includes the relative
    proportion of the crystals of the respective bin to the total amount of
    crystals.}
    \label{fig:histograms_rel}
    \end{figure}

    In Figure~\ref{fig:histograms_rel} we show the test classification error in
    each bin for the unit cell volume and the number of atoms in the asymmetric
    unit using the ResNet-101 model trained on diffractograms of synthetic
    crystals. The classification error is shown both for testing on
    diffractograms from synthetic crystals and on ICSD diffractograms. One can
    see that for small volumes and a small number of atoms in the asymmetric
    unit, the difference between classifying ICSD diffractograms and
    diffractograms from synthetic crystals is relatively small. As the volume
    and number of atoms in the asymmetric unit increase, the gap between the two
    errors increases, too. We already identified the uniformly sampled atom
    coordinates in the synthetic distribution as the main contributor to the gap
    in accuracy between the synthetic crystals and ICSD crystals. Therefore, it
    seems that the uniform sampling of atom coordinates works well for small 
    number of atoms in the asymmetric unit and small volumes, while the error 
    due to this sampling strategy increases slightly for higher volumes and higher
    number of atoms in the asymmetric unit.
    
    When looking at the distribution of crystals in the ICSD, the number of
    atoms in the asymmetric unit tends to be larger for lower-symmetry space
    groups (for example, in the triclinic crystal system) than for
    higher-symmetry space groups such as those from the cubic crystal system.
    Therefore, the increasing test error on diffractograms from ICSD crystals
    with a higher number of atoms in the asymmetric unit is especially relevant
    for these lower-symmetry space groups.
    It might be possible that a different scheme of generating atom positions in
    the unit cell (compared to the independent uniform sampling that we used)
    works better for a high number of atoms in the asymmetric unit.

    Overall, it is important to note that the distribution of ICSD crystals is (apart from a few Wyckoff position occupation probabilities which are exactly zero in the statistics dataset\footnote{Setting them to small non-zero values typically leads to the generation of rather large unit cells, as the general Wyckoff positions have high multiplicities.}) almost completely encompassed by the much larger distribution of snynthetic crystals that we used for training.
    However, due to finite training times and model capacity, a performance gap remains. This gap can be improved by using (substantially) more computing power or by narrowing the very general synthetic distribution, e.g., by using a different algorithm to generate atom positions.
    This indicates an inherent challenge in XRD classification but more generally in materials property prediction:
    Machine learning models are ultimately trained to be employed in real-world tasks, which are typically related to novel, i.e. yet unseen materials and structures.
    At the same time, the machine learning models are tested based on an IID assumption, i.e. the assumption that the distribution of training and testing data is the same.
    While not being a contradiction in the limit of infinite training data and model capacity, this becomes an (unsolvable) challenge in reality, when facing finite datasets and models.
    In our case, our model trained on a large distribution of synthetic crystal structures will likely generalize better to completely new crystal structures different from any crystal structure contained in the ICSD database.
    At the same time, it suffers from smaller ICSD test set errors, even though the ICSD distribution is contained in the synthetic data generation distribution.

\section{Conclusion}
We developed an algorithm based on the symmetry operations of the space groups
to generate synthetic crystals that follow the distribution found in the ICSD
database in terms of general descriptors like volume, density, or types of
elements. The generated crystals have randomly sampled coordinates and span a
wide range of structure types, many of which do not appear in the ICSD. We
showed that, compared to using ICSD crystals directly, simulating the training
data based on the synthetic crystals can improve the performance of tasks that
extract structural information from powder diffractograms, in this case, the
space group. \replaced{The more general dataset that also contains unstable structures helps to classify unseen stable crystal structures.}{Our results verify our hypothesis that crystals forming the
training dataset do not need to be stable or chemically viable.}

We trained on an infinite on-the-fly generated stream of synthetic crystals and simulated
batches of diffractograms using a distributed framework based on the
\emph{Python} library \emph{Ray}\supercite{moritzRayDistributedFramework2018a}.
This allows the training of very large networks without overfitting. The
best-performing model (ResNet-101) reached a space group classification accuracy
of 79.9\% vs. 56.1\% when training on ICSD structures directly. By performing
the train-test split using the structure type, we forced our models to not just
recognize structure types or individual structures, but to actually learn rules
to distinguish different space groups by their symmetry elements. This shows the
true generalization capabilities to new structure types and novel classes of
materials. We also demonstrated first steps toward applying the presented methodology to an experimental dataset. We expect further improvements in this area using improved models of experimental imperfections, as well as larger ML models and longer training times.

Even though models trained on the synthetic distribution transfer well when
tested on ICSD crystals, we found a gap of 12.3 percentage points (ResNet-101)
between the training accuracy on synthetic crystals and test accuracy on the
ICSD. We showed that the main contribution to this gap stems from the
independently uniformly sampled atom coordinates. An improved approach may be
needed to artificially generate more ordered structures, which contain more
ordered diffraction planes than a cloud of uniformly sampled points. This might
be especially important for crystals with a high number of atoms in the
asymmetric unit.

Lastly, the developed algorithm to synthetically generate crystals can be used
for other XRD-related tasks in the future, such as the extraction of crystallite sizes, lattice
parameters, information about the occupation of Wyckoff positions, etc.
Furthermore, instead of generating \replaced{synthetic}{completely random} crystals of all space
groups, one can also generate \deleted{random} crystals of given structure types to solve
more specialized tasks. This would allow the use of very large models for tasks
that are typically strongly limited by the dataset size when using only the entries of
the ICSD. Also, tasks concerning multi-phase diffractograms or augmentations such as
strain in given crystal structures can benefit from our batch-wise online
learning approach.
 
\subsection*{Data availability}
The source code of all machine learning models, of the generation of synthetic
crystals, of the optimized simulation of diffractograms, and of the distributed
computing architecture can be found on
\url{https://github.com/aimat-lab/ML4pXRDs} (v1.0). The used machine learning models are further
discussed in detail in the Supplementary Information.

The ICSD data used to evaluate the models \added{(database version 2021, June 15)} belongs to FIZ Karlsruhe, from which academic and non-academic licenses are available. \added{The RRUFF mineral database (access date: 2022, Jan 12) for the evaluation on experimental data can be obtained from \url{https://rruff.info/}.}


\subsection*{Author contributions}
All authors contributed to the idea and the preparation of the manuscript. 
H.S. implemented the methods and conducted the computational experiments.

\subsection*{Conflicts of interest}
There are no conflicts of interest to declare.

\subsection*{Acknowledgements}
P.F. acknowledges support by the Federal Ministry of Education and Research (BMBF) under Grant No. 01DM21001B (German-Canadian Materials Acceleration Center).
H.S. acknowledges financial support by the German Research Foundation (DFG) through the Research Training Group 2450 “Tailored Scale-Bridging Approaches to Computational Nanoscience”.
The authors acknowledge support by the state of Baden-Württemberg through bwHPC. Parts of this work were performed on the HoreKa supercomputer funded by the Ministry of Science, Research and the Arts Baden-Württemberg and by the Federal Ministry of Education and Research.

\printnomenclature

\clearpage



\bibliography{bib}

@misc{abadiTensorFlowLargeScaleMachine2015,
  title = {{{TensorFlow}}: {{Large-Scale Machine Learning}} on {{Heterogeneous Distributed Systems}}},
  shorttitle = {{{TensorFlow}}},
  author = {Abadi, Mart{\'i}n and Agarwal, Ashish and Barham, Paul and Brevdo, Eugene and Chen, Zhifeng and Citro, Craig and Corrado, Greg and Davis, Andy and Dean, Jeffrey and Devin, Matthieu and Ghemawat, Sanjay and Goodfellow, Ian and Harp, Andrew and Irving, Geoffrey and Isard, Michael and Jia, Yangqing and Jozefowicz, Rafal and Kaiser, Lukasz and Kudlur, Manjunath and Levenberg, Josh and Man{\'e}, Dan and Monga, Rajat and Moore, Sherry and Murray, Derek and Olah, Chris and Schuster, Mike and Shlens, Jonathon and Steiner, Benoit and Sutskever, Ilya and Talwar, Kunal and Tucker, Paul and Vanhoucke, Vincent and Vasudevan, Vijay and Vi{\'e}gas, Fernanda and Vinyals, Oriol and Warden, Pete and Wattenberg, Martin and Wicke, Martin and Yu, Yuan and Zheng, Xiaoqiang},
  year = {2015}
}

@article{bankoDeepLearningVisualization2021,
  title = {Deep Learning for Visualization and Novelty Detection in Large {{X-ray}} Diffraction Datasets},
  author = {Banko, Lars and Maffettone, Phillip M. and Naujoks, Dennis and Olds, Daniel and Ludwig, Alfred},
  year = {2021},
  month = jul,
  journal = {npj Comput. Mater.},
  volume = {7},
  number = {1},
  pages = {1--6},
  publisher = {{Nature Publishing Group}},
  issn = {2057-3960},
  doi = {10.1038/s41524-021-00575-9},
  urldate = {2021-09-30},
  copyright = {2021 The Author(s)},
  langid = {english}
}

@inproceedings{bergerhoff1987,
  booktitle = {Crystallographic {{Databases}}},
  author = {Bergerhoff, G. and Brown, I.D.},
  year = {1987},
  langid = {english}
}

@article{chakrabortyDeepCrystalStructure2022,
  title = {A Deep Crystal Structure Identification System for {{X-ray}} Diffraction Patterns},
  author = {Chakraborty, Abhik and Sharma, Raksha},
  year = {2022},
  month = apr,
  journal = {Vis. Comput.},
  volume = {38},
  number = {4},
  pages = {1275--1282},
  issn = {1432-2315},
  doi = {10.1007/s00371-021-02165-8},
  urldate = {2022-09-13},
  langid = {english}
}

@article{chitturiAutomatedPredictionLattice2021,
  title = {Automated Prediction of Lattice Parameters from {{X-ray}} Powder Diffraction Patterns},
  author = {Chitturi, S. R. and Ratner, D. and Walroth, R. C. and Thampy, V. and Reed, E. J. and Dunne, M. and Tassone, C. J. and Stone, K. H.},
  year = {2021},
  month = dec,
  journal = {J. Appl. Crystallogr.},
  volume = {54},
  number = {6},
  pages = {1799--1810},
  publisher = {{International Union of Crystallography}},
  issn = {1600-5767},
  doi = {10.1107/S1600576721010840},
  urldate = {2021-12-22},
  copyright = {https://creativecommons.org/licenses/by/4.0/},
  langid = {english}
}

@article{dongDeepConvolutionalNeural2021,
  title = {A Deep Convolutional Neural Network for Real-Time Full Profile Analysis of Big Powder Diffraction Data},
  author = {Dong, Hongyang and Butler, Keith T. and Matras, Dorota and Price, Stephen W. T. and Odarchenko, Yaroslav and Khatry, Rahul and Thompson, Andrew and Middelkoop, Vesna and Jacques, Simon D. M. and Beale, Andrew M. and Vamvakeros, Antonis},
  year = {2021},
  month = may,
  journal = {npj Comput. Mater.},
  volume = {7},
  number = {1},
  pages = {1--9},
  publisher = {{Nature Publishing Group}},
  issn = {2057-3960},
  doi = {10.1038/s41524-021-00542-4},
  urldate = {2022-06-01},
  copyright = {2021 The Author(s)},
  langid = {english}
}

@article{fengMethodArtificialIntelligence2019,
  title = {Method of Artificial Intelligence Algorithm to Improve the Automation Level of {{Rietveld}} Refinement},
  author = {Feng, Zhenjie and Hou, Qiang and Zheng, Yonglei and Ren, Wei and Ge, Jun-Yi and Li, Tao and Cheng, Cheng and Lu, Wencong and Cao, Shixun and Zhang, Jincang and Zhang, Tongyi},
  year = {2019},
  month = jan,
  journal = {Comput. Mater. Sci.},
  volume = {156},
  pages = {310--314},
  issn = {0927-0256},
  doi = {10.1016/j.commatsci.2018.10.006},
  urldate = {2021-09-13},
  langid = {english}
}

@misc{francoisKeras,
  title = {Keras},
  author = {Fran{\c c}ois, Chollet}
}

@article{fredericksPyXtalPythonLibrary2021,
  title = {{{PyXtal}}: {{A Python}} Library for Crystal Structure Generation and Symmetry Analysis},
  shorttitle = {{{PyXtal}}},
  author = {Fredericks, Scott and Parrish, Kevin and Sayre, Dean and Zhu, Qiang},
  year = {2021},
  month = apr,
  journal = {Comput. Phys. Commun.},
  volume = {261},
  pages = {107810},
  issn = {0010-4655},
  doi = {10.1016/j.cpc.2020.107810},
  urldate = {2022-06-27},
  langid = {english}
}

@book{gilmoreInternationalTablesCrystallography2019,
  title = {International {{Tables}} for {{Crystallography Volume H}}: {{Powder}} Diffraction},
  shorttitle = {International {{Tables}} for {{Crystallography Volume H}}},
  editor = {Gilmore, C. J. and Kaduk, J. A. and Schenk, H.},
  year = {2019},
  month = sep,
  series = {{{IUCr Series}}. {{International Tables}} of {{Crystallography}}},
  edition = {1},
  volume = {H},
  publisher = {{Wiley}},
  urldate = {2021-10-06},
  isbn = {978-1-118-41628-0}
}

@book{goodfellowDeepLearning2016,
  title = {Deep {{Learning}}},
  author = {Goodfellow, Ian and Bengio, Yoshua and Courville, Aaron},
  year = {2016},
  publisher = {{The MIT Press}},
  langid = {english}
}

@article{habershonPowderDiffractionIndexing2004,
  title = {Powder {{Diffraction Indexing}} as a {{Pattern Recognition Problem}}:\, {{A New Approach}} for {{Unit Cell Determination Based}} on an {{Artificial Neural Network}}},
  shorttitle = {Powder {{Diffraction Indexing}} as a {{Pattern Recognition Problem}}},
  author = {Habershon, Scott and Cheung, Eugene Y. and Harris, Kenneth D. M. and Johnston, Roy L.},
  year = {2004},
  month = feb,
  journal = {J. Phys. Chem. A},
  volume = {108},
  number = {5},
  pages = {711--716},
  publisher = {{American Chemical Society}},
  issn = {1089-5639},
  doi = {10.1021/jp0310596},
  urldate = {2021-10-01}
}

@article{harrisContemporaryAdvancesUse2001,
  title = {Contemporary {{Advances}} in the {{Use}} of {{Powder X-Ray Diffraction}} for {{Structure Determination}}},
  author = {Harris, Kenneth D. M. and Tremayne, Maryjane and Kariuki, Benson M.},
  year = {2001},
  journal = {Angew. Chem., Int. Ed.},
  volume = {40},
  number = {9},
  pages = {1626--1651},
  issn = {1521-3773},
  doi = {10.1002/1521-3773(20010504)40:9{$<$}1626::AID-ANIE16260{$>$}3.0.CO;2-7},
  urldate = {2023-03-22},
  langid = {english}
}

@inproceedings{heDeepResidualLearning2016,
  title = {Deep {{Residual Learning}} for {{Image Recognition}}},
  booktitle = {2016 {{IEEE Conference}} on {{Computer Vision}} and {{Pattern Recognition}} ({{CVPR}})},
  author = {He, Kaiming and Zhang, Xiangyu and Ren, Shaoqing and Sun, Jian},
  year = {2016},
  month = jun,
  pages = {770--778},
  issn = {1063-6919},
  doi = {10.1109/CVPR.2016.90}
}

@article{holderTutorialPowderXray2019,
  title = {Tutorial on {{Powder X-ray Diffraction}} for {{Characterizing Nanoscale Materials}}},
  author = {Holder, Cameron F. and Schaak, Raymond E.},
  year = {2019},
  month = jul,
  journal = {ACS Nano},
  volume = {13},
  number = {7},
  pages = {7359--7365},
  publisher = {{American Chemical Society}},
  issn = {1936-0851},
  doi = {10.1021/acsnano.9b05157},
  urldate = {2023-03-22}
}

@article{kadukTypicalValuesRietveld2011,
  title = {Typical Values of {{Rietveld}} Instrument Profile Coefficients},
  author = {Kaduk, James and Reid, Joel},
  year = {2011},
  month = mar,
  journal = {Powder Diffraction - POWDER DIFFR},
  volume = {26},
  pages = {88--94},
  doi = {10.1154/1.3548128}
}

@misc{kingmaAdamMethodStochastic2017,
  title = {Adam: {{A Method}} for {{Stochastic Optimization}}},
  shorttitle = {Adam},
  author = {Kingma, Diederik P. and Ba, Jimmy},
  year = {2017},
  month = jan,
  number = {arXiv:1412.6980},
  eprint = {1412.6980},
  publisher = {{arXiv}},
  doi = {10.48550/arXiv.1412.6980},
  urldate = {2022-06-28},
  archiveprefix = {arxiv}
}

@book{lafuentePowerDatabasesRRUFF2015,
  title = {The Power of Databases: {{The RRUFF}} Project in {{Highlights}} in {{Mineralogical Crystallography}}},
  author = {Lafuente, Barbara and Downs, R. T. and Yang, H. and Stone, N.},
  year = {2015},
  month = nov,
  publisher = {{De Gruyter (O)}},
  isbn = {978-3-11-041710-4},
  langid = {english}
}

@inproceedings{lamNumbaLLVMbasedPython2015,
  title = {Numba: A {{LLVM-based Python JIT}} Compiler},
  shorttitle = {Numba},
  booktitle = {Proceedings of the {{Second Workshop}} on the {{LLVM Compiler Infrastructure}} in {{HPC}}},
  author = {Lam, Siu Kwan and Pitrou, Antoine and Seibert, Stanley},
  year = {2015},
  month = nov,
  series = {{{LLVM}} '15},
  pages = {1--6},
  publisher = {{Association for Computing Machinery}},
  address = {{New York, NY, USA}},
  doi = {10.1145/2833157.2833162},
  urldate = {2022-09-13},
  isbn = {978-1-4503-4005-2}
}

@article{leeDatadrivenXRDAnalysis2021,
  title = {A Data-Driven {{XRD}} Analysis Protocol for Phase Identification and Phase-Fraction Prediction of Multiphase Inorganic Compounds},
  author = {Lee, Jin-Woong and Bae~Park, Woon and Kim, Minseuk and Singh, Satendra Pal and Pyo, Myoungho and Sohn, Kee-Sun},
  year = {2021},
  journal = {Inorg. Chem. Front.},
  volume = {8},
  number = {10},
  pages = {2492--2504},
  publisher = {{Royal Society of Chemistry}},
  doi = {10.1039/D0QI01513J},
  urldate = {2021-09-10},
  langid = {english}
}

@article{leeDeeplearningTechniquePhase2020,
  title = {A Deep-Learning Technique for Phase Identification in Multiphase Inorganic Compounds Using Synthetic {{XRD}} Powder Patterns},
  author = {Lee, Jin-Woong and Park, Woon Bae and Lee, Jin Hee and Singh, Satendra Pal and Sohn, Kee-Sun},
  year = {2020},
  month = jan,
  journal = {Nat. Commun.},
  volume = {11},
  number = {1},
  pages = {86},
  publisher = {{Nature Publishing Group}},
  issn = {2041-1723},
  doi = {10.1038/s41467-019-13749-3},
  urldate = {2021-09-06},
  copyright = {2020 The Author(s)},
  langid = {english}
}

@article{maffettoneCrystallographyCompanionAgent2021,
  title = {Crystallography Companion Agent for High-Throughput Materials Discovery},
  author = {Maffettone, Phillip M. and Banko, Lars and Cui, Peng and Lysogorskiy, Yury and Little, Marc A. and Olds, Daniel and Ludwig, Alfred and Cooper, Andrew I.},
  year = {2021},
  month = apr,
  journal = {Nat. Comput. Sci.},
  volume = {1},
  number = {4},
  pages = {290--297},
  publisher = {{Nature Publishing Group}},
  issn = {2662-8457},
  doi = {10.1038/s43588-021-00059-2},
  urldate = {2021-09-30},
  copyright = {2021 This is a U.S. government work and not under copyright protection in the U.S.; foreign copyright protection may apply},
  langid = {english}
}

@inproceedings{moritzRayDistributedFramework2018a,
  title = {Ray: {{A Distributed Framework}} for {{Emerging AI Applications}}},
  shorttitle = {Ray},
  booktitle = {13th {{USENIX Symposium}} on {{Operating Systems Design}} and {{Implementation}} ({{OSDI}} 18)},
  author = {Moritz, Philipp and Nishihara, Robert and Wang, Stephanie and Tumanov, Alexey and Liaw, Richard and Liang, Eric and Elibol, Melih and Yang, Zongheng and Paul, William and Jordan, Michael I. and Stoica, Ion},
  year = {2018},
  pages = {561--577},
  urldate = {2022-09-13},
  isbn = {978-1-939133-08-3},
  langid = {english}
}

@article{ongPythonMaterialsGenomics2013,
  title = {Python {{Materials Genomics}} (Pymatgen): {{A}} Robust, Open-Source Python Library for Materials Analysis},
  shorttitle = {Python {{Materials Genomics}} (Pymatgen)},
  author = {Ong, Shyue Ping and Richards, William Davidson and Jain, Anubhav and Hautier, Geoffroy and Kocher, Michael and Cholia, Shreyas and Gunter, Dan and Chevrier, Vincent L. and Persson, Kristin A. and Ceder, Gerbrand},
  year = {2013},
  month = feb,
  journal = {Comput. Mater. Sci.},
  volume = {68},
  pages = {314--319},
  issn = {0927-0256},
  doi = {10.1016/j.commatsci.2012.10.028},
  urldate = {2022-06-21},
  langid = {english}
}

@article{oviedoFastInterpretableClassification2019,
  title = {Fast and Interpretable Classification of Small {{X-ray}} Diffraction Datasets Using Data Augmentation and Deep Neural Networks},
  author = {Oviedo, Felipe and Ren, Zekun and Sun, Shijing and Settens, Charles and Liu, Zhe and Hartono, Noor Titan Putri and Ramasamy, Savitha and DeCost, Brian L. and Tian, Siyu I. P. and Romano, Giuseppe and Gilad Kusne, Aaron and Buonassisi, Tonio},
  year = {2019},
  month = may,
  journal = {npj Comput. Mater.},
  volume = {5},
  number = {1},
  pages = {1--9},
  publisher = {{Nature Publishing Group}},
  issn = {2057-3960},
  doi = {10.1038/s41524-019-0196-x},
  urldate = {2021-09-16},
  copyright = {2019 The Author(s)},
  langid = {english}
}

@article{ozakiAutomatedCrystalStructure2020,
  title = {Automated Crystal Structure Analysis Based on Blackbox Optimisation},
  author = {Ozaki, Yoshihiko and Suzuki, Yuta and Hawai, Takafumi and Saito, Kotaro and Onishi, Masaki and Ono, Kanta},
  year = {2020},
  month = jun,
  journal = {npj Comput. Mater.},
  volume = {6},
  number = {1},
  pages = {1--7},
  publisher = {{Nature Publishing Group}},
  issn = {2057-3960},
  doi = {10.1038/s41524-020-0330-9},
  urldate = {2021-09-17},
  copyright = {2020 The Author(s)},
  langid = {english}
}

@article{parkClassificationCrystalStructure2017,
  title = {Classification of Crystal Structure Using a Convolutional Neural Network},
  author = {Park, W. B. and Chung, J. and Jung, J. and Sohn, K. and Singh, S. P. and Pyo, M. and Shin, N. and Sohn, K.-S.},
  year = {2017},
  month = jul,
  journal = {IUCrJ},
  volume = {4},
  number = {4},
  pages = {486--494},
  publisher = {{International Union of Crystallography}},
  issn = {2052-2525},
  doi = {10.1107/S205225251700714X},
  urldate = {2021-09-06},
  copyright = {http://creativecommons.org/licenses/by/2.0/uk},
  langid = {english}
}

@article{radovicMachineLearningEnergy2018,
  title = {Machine Learning at the Energy and Intensity Frontiers of Particle Physics},
  author = {Radovic, Alexander and Williams, Mike and Rousseau, David and Kagan, Michael and Bonacorsi, Daniele and Himmel, Alexander and Aurisano, Adam and Terao, Kazuhiro and Wongjirad, Taritree},
  year = {2018},
  month = aug,
  journal = {Nature},
  volume = {560},
  number = {7716},
  pages = {41--48},
  publisher = {{Nature Publishing Group}},
  issn = {1476-4687},
  doi = {10.1038/s41586-018-0361-2},
  urldate = {2023-03-22},
  copyright = {2018 Springer Nature Limited},
  langid = {english}
}

@article{rahmanianEnablingModularAutonomous2022,
  title = {Enabling {{Modular Autonomous Feedback-Loops}} in {{Materials Science}} through {{Hierarchical Experimental Laboratory Automation}} and {{Orchestration}}},
  author = {Rahmanian, Fuzhan and Flowers, Jackson and Guevarra, Dan and Richter, Matthias and Fichtner, Maximilian and Donnely, Phillip and Gregoire, John M. and Stein, Helge S.},
  year = {2022},
  journal = {Adv. Mater. Interfaces},
  volume = {9},
  number = {8},
  pages = {2101987},
  issn = {2196-7350},
  doi = {10.1002/admi.202101987},
  urldate = {2022-06-24},
  langid = {english}
}

@article{schuetzkeEnhancingDeeplearningTraining2021,
  title = {Enhancing Deep-Learning Training for Phase Identification in Powder {{X-ray}} Diffractograms},
  author = {Schuetzke, J. and Benedix, A. and Mikut, R. and Reischl, M.},
  year = {2021},
  month = may,
  journal = {IUCrJ},
  volume = {8},
  number = {3},
  pages = {408--420},
  publisher = {{International Union of Crystallography}},
  issn = {2052-2525},
  doi = {10.1107/S2052252521002402},
  urldate = {2021-09-16},
  copyright = {https://creativecommons.org/licenses/by/4.0/},
  langid = {english}
}

@misc{schuetzkeUniversalSyntheticDataset2022a,
  title = {A Universal Synthetic Dataset for Machine Learning on Spectroscopic Data},
  author = {Schuetzke, Jan and Szymanski, Nathan J. and Reischl, Markus},
  year = {2022},
  month = jun,
  number = {arXiv:2206.06031},
  eprint = {2206.06031},
  primaryclass = {cond-mat},
  publisher = {{arXiv}},
  doi = {10.48550/arXiv.2206.06031},
  urldate = {2023-03-24},
  archiveprefix = {arxiv}
}

@article{seaboldStatsmodelsEconometricStatistical2010,
  title = {Statsmodels: {{Econometric}} and {{Statistical Modeling}} with {{Python}}},
  shorttitle = {Statsmodels},
  author = {Seabold, Skipper and Perktold, Josef},
  year = {2010},
  month = jan,
  journal = {Proceedings of the 9th Python in Science Conference},
  volume = {2010}
}

@article{sunClassificationImbalancedData2009,
  title = {Classification of Imbalanced Data: A Review},
  shorttitle = {Classification of Imbalanced Data},
  author = {Sun, Yanmin and Wong, Andrew K. C. and Kamel, Mohamed S.},
  year = {2009},
  month = jun,
  journal = {Int. J. Patt. Recogn. Artif. Intell.},
  volume = {23},
  number = {04},
  pages = {687--719},
  publisher = {{World Scientific Publishing Co.}},
  issn = {0218-0014},
  doi = {10.1142/S0218001409007326},
  urldate = {2022-06-24}
}

@article{suzukiSymmetryPredictionKnowledge2020,
  title = {Symmetry Prediction and Knowledge Discovery from {{X-ray}} Diffraction Patterns Using an Interpretable Machine Learning Approach},
  author = {Suzuki, Yuta and Hino, Hideitsu and Hawai, Takafumi and Saito, Kotaro and Kotsugi, Masato and Ono, Kanta},
  year = {2020},
  month = dec,
  journal = {Sci. Rep.},
  volume = {10},
  number = {1},
  pages = {21790},
  publisher = {{Nature Publishing Group}},
  issn = {2045-2322},
  doi = {10.1038/s41598-020-77474-4},
  urldate = {2021-09-26},
  copyright = {2020 The Author(s)},
  langid = {english}
}

@article{szymanskiProbabilisticDeepLearning2021,
  title = {Probabilistic {{Deep Learning Approach}} to {{Automate}} the {{Interpretation}} of {{Multi-phase Diffraction Spectra}}},
  author = {Szymanski, Nathan J. and Bartel, Christopher J. and Zeng, Yan and Tu, Qingsong and Ceder, Gerbrand},
  year = {2021},
  month = jun,
  journal = {Chem. Mater.},
  volume = {33},
  number = {11},
  pages = {4204--4215},
  publisher = {{American Chemical Society}},
  issn = {0897-4756},
  doi = {10.1021/acs.chemmater.1c01071},
  urldate = {2021-09-10}
}

@misc{togoSpglibSoftwareLibrary2018,
  title = {Spglib: A Software Library for Crystal Symmetry Search},
  shorttitle = {\$\textbackslash texttt\{\vphantom\}{{Spglib}}\vphantom\{\}\$},
  author = {Togo, Atsushi and Tanaka, Isao},
  year = {2018},
  month = aug,
  number = {arXiv:1808.01590},
  eprint = {1808.01590},
  publisher = {{arXiv}},
  doi = {10.48550/arXiv.1808.01590},
  urldate = {2022-06-28},
  archiveprefix = {arxiv}
}

@article{vecseiNeuralNetworkBased2019,
  title = {Neural Network Based Classification of Crystal Symmetries from X-Ray Diffraction Patterns},
  author = {Vecsei, Pascal Marc and Choo, Kenny and Chang, Johan and Neupert, Titus},
  year = {2019},
  month = jun,
  journal = {Phys. Rev. B},
  volume = {99},
  number = {24},
  pages = {245120},
  publisher = {{American Physical Society}},
  doi = {10.1103/PhysRevB.99.245120},
  urldate = {2021-09-06}
}

@article{virtanenSciPyFundamentalAlgorithms2020,
  title = {{{SciPy}} 1.0: Fundamental Algorithms for Scientific Computing in {{Python}}},
  shorttitle = {{{SciPy}} 1.0},
  author = {Virtanen, Pauli and Gommers, Ralf and Oliphant, Travis E. and Haberland, Matt and Reddy, Tyler and Cournapeau, David and Burovski, Evgeni and Peterson, Pearu and Weckesser, Warren and Bright, Jonathan and {van der Walt}, St{\'e}fan J. and Brett, Matthew and Wilson, Joshua and Millman, K. Jarrod and Mayorov, Nikolay and Nelson, Andrew R. J. and Jones, Eric and Kern, Robert and Larson, Eric and Carey, C. J. and Polat, {\.I}lhan and Feng, Yu and Moore, Eric W. and VanderPlas, Jake and Laxalde, Denis and Perktold, Josef and Cimrman, Robert and Henriksen, Ian and Quintero, E. A. and Harris, Charles R. and Archibald, Anne M. and Ribeiro, Ant{\^o}nio H. and Pedregosa, Fabian and {van Mulbregt}, Paul},
  year = {2020},
  month = mar,
  journal = {Nat Methods},
  volume = {17},
  number = {3},
  pages = {261--272},
  publisher = {{Nature Publishing Group}},
  issn = {1548-7105},
  doi = {10.1038/s41592-019-0686-2},
  urldate = {2022-08-31},
  copyright = {2020 The Author(s)},
  langid = {english}
}

@article{wangRapidIdentificationXray2020,
  title = {Rapid {{Identification}} of {{X-ray Diffraction Patterns Based}} on {{Very Limited Data}} by {{Interpretable Convolutional Neural Networks}}},
  author = {Wang, Hong and Xie, Yunchao and Li, Dawei and Deng, Heng and Zhao, Yunxin and Xin, Ming and Lin, Jian},
  year = {2020},
  month = apr,
  journal = {J. Chem. Inf. Model.},
  volume = {60},
  number = {4},
  pages = {2004--2011},
  publisher = {{American Chemical Society}},
  issn = {1549-9596},
  doi = {10.1021/acs.jcim.0c00020},
  urldate = {2021-09-29}
}

@inproceedings{wuGroupNormalization2018,
  title = {Group {{Normalization}}},
  booktitle = {Proceedings of the {{European Conference}} on {{Computer Vision}} ({{ECCV}})},
  author = {Wu, Yuxin and He, Kaiming},
  year = {2018},
  pages = {3--19},
  doi = {10.1007/s11263-019-01198-w}
}

@misc{yuTensorFlowModelGarden,
  title = {{{TensorFlow Model Garden}}},
  author = {Yu, Hongkun and Chen, Chen and Xianzhi, Du and Yeqing, Li and Abdullah, Rashwan and Le, Hou and Pengchong, Jin and Fan, Yang and Frederick, Liu and Jaeyoun, Kim and Jing, Li},
  doi = {10.5281/zenodo.11813},
  urldate = {2022-08-18}
}

@article{zalogaCrystalSymmetryClassification2020,
  title = {Crystal Symmetry Classification from Powder {{X-ray}} Diffraction Patterns Using a Convolutional Neural Network},
  author = {Zaloga, Alexander N. and Stanovov, Vladimir V. and Bezrukova, Oksana E. and Dubinin, Petr S. and Yakimov, Igor S.},
  year = {2020},
  month = dec,
  journal = {Mater. Today Commun.},
  volume = {25},
  pages = {101662},
  issn = {2352-4928},
  doi = {10.1016/j.mtcomm.2020.101662},
  urldate = {2022-04-06},
  langid = {english}
}
\bibliographystyle{rsc} 

\end{document}


\maketitle

\section{Generating synthetic crystals} \label{sec:generating_crystals} 

Here we describe the algorithm to generate synthetic crystals in more detail.
To generate a single crystal, the following steps are executed:

\begin{enumerate}
    \itemsep0em
    \item \replaced{Sampling of a space group from the space group distribution of the ICSD.}{Random selection of a space group. We follow the space group distribution of the ICSD to allow comparison with previous work (see Section~2.1 of the paper).}
    \item The number of unique elements in the crystal is drawn from a discrete
    distribution extracted from the crystals in the ICSD belonging to the
    specified space group.
    \item The unique elements are drawn, also from a discrete probability
    distribution from the crystals in the ICSD belonging to the specified space
    group.
    \item For each of the unique elements, the number of repetitions in the
    asymmetric unit is chosen, and for each repetition, a Wyckoff position is
    \deleted{randomly} selected. Again, both the probability of the number of repetitions and the Wyckoff
    occupation probabilities are extracted from the ICSD for the specified space
    group. We do not place more than one atom onto a Wyckoff position that does not
    have a degree of freedom.
    \item For each atom placed on a Wyckoff position, uniformly distributed random fractional
    coordinates are drawn.
    \item Lattice parameters (normalized to unity volume) of the crystal system
    that the specified space group belongs to are drawn from a kernel density
    estimate \deleted{(KDE)}
    of the ICSD. The bandwidth is chosen based on Scott's rule
    (see the \emph{SciPy} \supercite{virtanenSciPyFundamentalAlgorithms2020}
    implementation of \replaced{the kernel density estimate}{KDE}). 
    \item We generated a \replaced{kernel density estimate}{KDE} of the volume conditioned\footnote{For this
        conditional kernel density estimate, we used the implementation of
        \emph{statsmodels} \supercite{seaboldStatsmodelsEconometricStatistical2010}
        with the normal reference rule of thumb to estimate the bandwidth.} on \\ $
        \sum_i 4/3 \pi \left(\frac{r_{i\text{;cov}} +
        r_{i\text{;VdW}}}{2}\right)^3 = V_\text{atomic} $, where the sum covers
        all atoms in the conventional unit cell, $r_{i\text{;cov}}$ is the
        atomic covalent radius, and $r_{i\text{;VdW}}$ is the atomic van der
        Waals radius. The \replaced{kernel density estimate}{KDE} was generated from all crystals of the ICSD
        belonging to the specified space group. Then, $ V_\text{atomic} $ is
        calculated for the chosen atoms in the conventional unit cell and the volume is
        drawn based on the \replaced{kernel density estimate}{KDE} conditioned on $V_\text{atomic}$. The lattice parameters (chosen in
        the previous step) are further scaled by the cube root of the chosen volume.
    \item Space group symmetry operations are applied using \emph{Python}
    library \emph{PyXtal} \supercite{fredericksPyXtalPythonLibrary2021}.
\end{enumerate} 

When generating a crystal of a specific space group without placing an atom
on the general Wyckoff position, it is not always the case that the crystal
belongs to that space group. To prevent wrong space group assignments, we use the \emph{Pymatgen}
\supercite{ongPythonMaterialsGenomics2013} interface to \emph{Spglib}
\supercite{togoSpglibSoftwareLibrary2018} to check the space group of each
crystal after its generation. 
If the space group deviates, we generate a new
crystal with the same number of unique elements as before, in order to not 
distort the distribution of number of unique elements extracted from the ICSD.
If the generation
fails 20 times in total, we start from the beginning with a new number of
unique elements. 
    
\section{Machine learning} \label{sec:ml_methods} 

    \subsection{Models}
    We now want to describe the machine learning models that we used for the
    classification of space groups in more detail. Powder diffractograms include
    similar features (peaks) at different locations and the position of a
    feature in the diffractogram has a spatial meaning. This suggests that the
    properties of the convolution operation, namely the parameter sharing (with
    sparse connectivity) and equivariance\supercite{goodfellowDeepLearning2016},
    might be beneficial when processing powder diffractograms.

    As a baseline, we first used the two CNN architectures used by
    \citeauthor{parkClassificationCrystalStructure2017}
    \supercite{parkClassificationCrystalStructure2017} for the classification of
    extinction groups and space groups. Since our training dataset is an
    infinite stream of diffractograms and we do not have to worry about
    overfitting, we further used the deeper architectures ResNet-10, ResNet-50, and
    ResNet-101\supercite{heDeepResidualLearning2016}. All architectures are now
    described in detail.

    \subsubsection*{Architectures by \citeauthor{parkClassificationCrystalStructure2017}}

    \citeauthor{parkClassificationCrystalStructure2017}\supercite{parkClassificationCrystalStructure2017}
    introduced three models, one for the classification of the crystal system,
    one for extinction groups, and one for space groups. We used only the last
    two models and call them ``parkCNN medium'' and ``parkCNN big'',
    respectively.

\begin{figure}[!htb]
    \centering
    \includegraphics{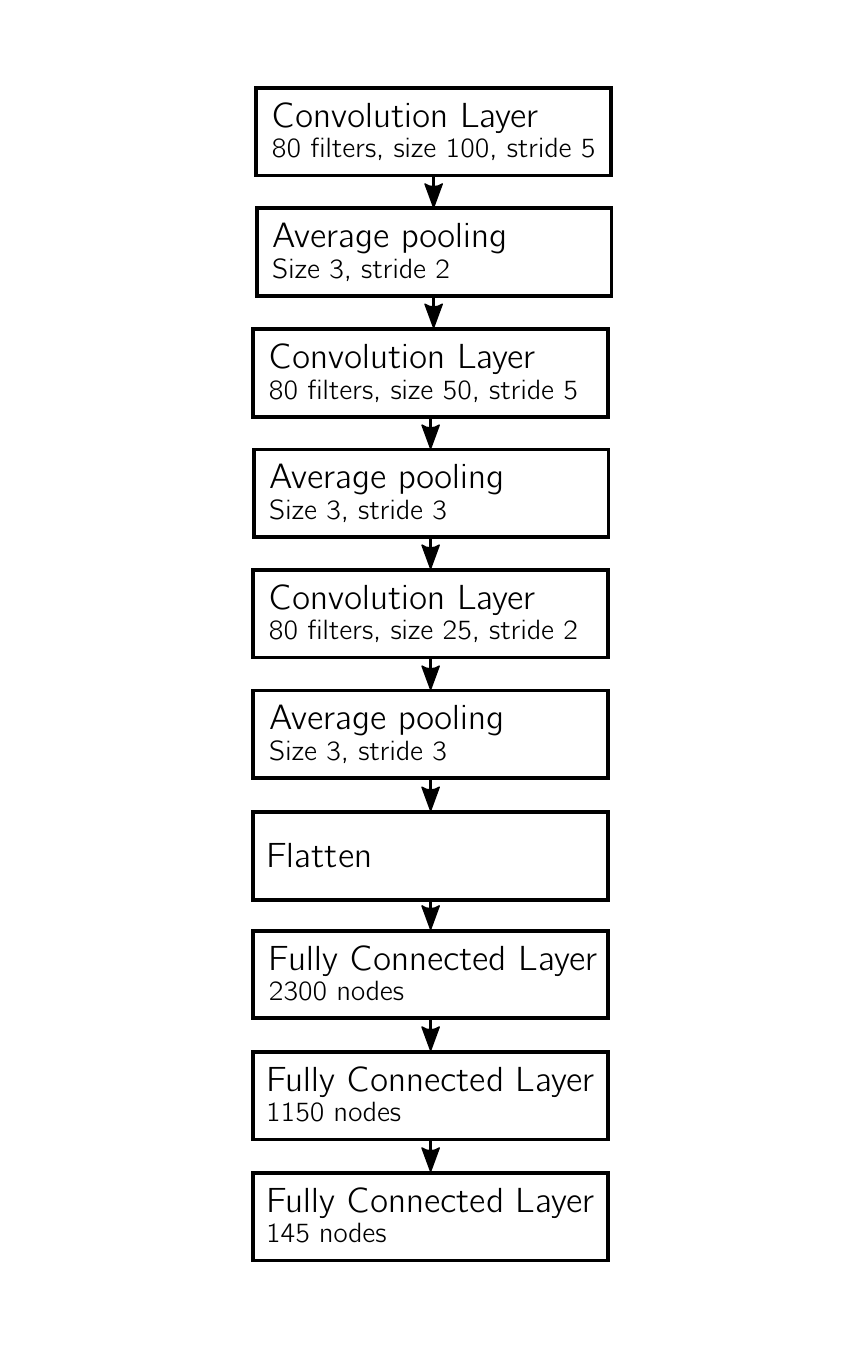}
    \caption{The CNN architecture for space group classification (``parkCNN
    big'') as introduced by \citeauthor{parkClassificationCrystalStructure2017}
    \supercite{parkClassificationCrystalStructure2017}. Each convolution or
    fully connected layer is implicitly followed by a ReLU activation, the
    output uses a softmax activation. We used only 145 target space groups
    instead of 230, since the remaining space groups did not have enough entries
    present in the ICSD to extract enough statistics for the synthetic
    generation of crystals. Furthermore,
    \citeauthor{parkClassificationCrystalStructure2017} used an input length of
    10001 instead of our input length of 8501. A dropout rate of 30\% is used
    after the activations of the convolution blocks. Dropout with a rate of 50\%
    is used after the activations of the fully connected layers. However,
    dropout is only used if the model is directly trained on ICSD
    diffractograms, not when using the synthetic data.}
    \label{fig:park_architecture}
    \end{figure}

    ``parkCNN big'' consists of three convolution layers with average pooling,
    two hidden fully connected layers with 2300 and 1150 nodes, and a
    145-dimensional softmax output. The architecture is summarized in
    Figure~\ref{fig:park_architecture}. The ``parkCNN medium'' model has fewer
    parameters than ``parkCNN big'' since the two hidden fully connected layers
    have 4040 and 202 nodes.

    \subsubsection*{ResNet architecture}
    With increasing size of the training dataset and increasing difficulty
    of the chosen task, the number of model parameters needs to be
    increased, too.

    In principle, a deeper model with additional layers should always be able to
    express the same solution of a shallower model by simply ``learning'' an
    identity map in addition to the shallower model. In practice, however, a
    degradation problem for CNNs with increasing depth has been observed and
    very deep models can perform worse than their shallow counterpart
    \supercite{heDeepResidualLearning2016}. Therefore, the ResNet architecture
    developed by \citeauthor{heDeepResidualLearning2016}
    \supercite{heDeepResidualLearning2016} at Microsoft in 2015 introduced additional
    skip connections, where information is able to simply flow past the
    convolution layer and is added to its output. This makes it possible for
    needed information of the input or earlier layers to flow further into the
    model without degradation.

    \begin{figure}[!htb]
    \centering
    \includegraphics{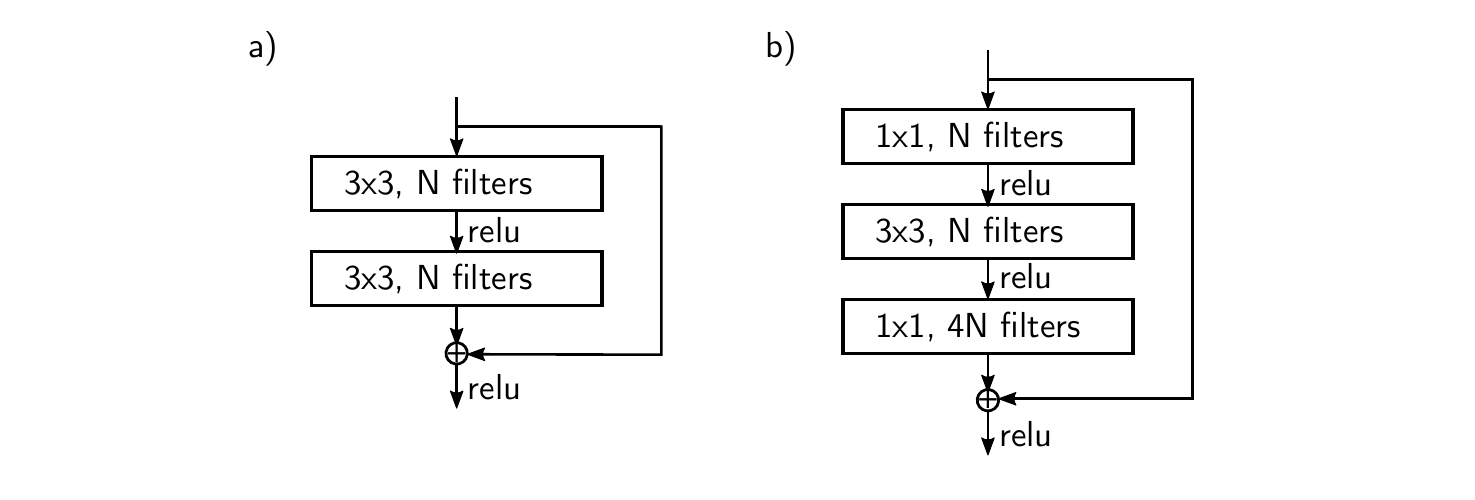}
    \caption{a) ResNet residual block, the main building block of the shallower
    variants of the ResNet architecture. b) ResNet bottleneck block, the main
    building block of the deeper variants of the ResNet architecture. All
    convolution operations are implicitly followed by a batch normalization
    layer. In both cases, a skip connection allows information to directly flow
    past the convolution operations. (Illustration based on
    \supercite{heDeepResidualLearning2016})} 
    \label{fig:resnet_bottleneck}
    \end{figure}

    Figure~\ref{fig:resnet_bottleneck}a visualizes the residual block used for
    the shallower versions of the ResNet architecture (up to 34 layers). Figure~\ref{fig:resnet_bottleneck}b visualizes the bottleneck block used for the
    deeper variants (50 and more layers). This type of building block is called
    a bottleneck block since it first reduces the number of channels using a 1x1
    convolution operation with N filters. Then, the main convolution with a
    3x3 kernel is performed, followed by a third 1x1 convolution that upscales
    to 4N channels. This down- and upscaling of the number of channels is
    performed to increase the performance of the model. All convolution
    operations of both types of building blocks are followed by batch
    normalization implicitly.
    
    In the simplest case, the skip connection of the residual and bottleneck
    block is simply an identity mapping and added to the output of the block.
    However, if the number of input channels and dimensions of a block are
    different from those in the output, a projection in the form of a 1x1
    convolution with the necessary number of filters and stride (usually 2) is
    used instead of the identity.

    \begin{table}[!htb]
        \centering
        \caption{Composition of the ResNet-10, ResNet-50, and ResNet-101
        architectures \supercite{heDeepResidualLearning2016}. The architectures
        are to be read from top to bottom. Square brackets indicate a residual
        or bottleneck building block with the respective number of filters for
        each convolution.}
        \vspace*{2mm}
        \begin{tabular}{ccccc}
            output size & layer name & ResNet-10 & ResNet-50 & ResNet-101 \\
            \toprule
            112x112 & initial conv. & \multicolumn{3}{c}{$ 7 \times 7 $ conv., 64 channels, stride 2} \\
            \midrule
            56x56 & initial pool & \multicolumn{3}{c}{$3 \times 3$ max pool, stride 2} \\
            \midrule
            56x56 & block group 1 & $ \left[\begin{array}{c} 3 \times 3,64 \\ 3 \times 3,64 \end{array}\right] $ $\times 1$ & $ \left[\begin{array}{c} 1 \times 1,64 \\ 3 \times 3,64 \\ 1 \times 1,256 \end{array}\right] $ $\times 3$ & $ \left[\begin{array}{c} 1 \times 1,64 \\ 3 \times 3,64 \\ 1 \times 1,256 \end{array}\right] $ $\times 3$ \\[5mm]
            \midrule
            28x28 & block group 2 & $ \left[\begin{array}{c} 3 \times 3,128 \\ 3 \times 3,128 \end{array}\right] $ $\times 1$ & $ \left[\begin{array}{c} 1 \times 1,128 \\ 3 \times 3,128 \\ 1 \times 1,512 \end{array}\right] $ $\times 4$ & $ \left[\begin{array}{c} 1 \times 1,128 \\ 3 \times 3,128 \\ 1 \times 1,512 \end{array}\right] $ $\times 4$ \\[5mm]
            \midrule
            14x14 & block group 3 & $ \left[\begin{array}{c} 3 \times 3,256 \\ 3 \times 3,256 \end{array}\right] $ $\times 1$ & $ \left[\begin{array}{c} 1 \times 1,256 \\ 3 \times 3,256 \\ 1 \times 1,1024 \end{array}\right] $ $\times 6$ & $ \left[\begin{array}{c} 1 \times 1,256 \\ 3 \times 3,256 \\ 1 \times 1,1024 \end{array}\right] $ $\times 23$\\[5mm]
            \midrule
            7x7 & block group 4 & $ \left[\begin{array}{c} 3 \times 3,512 \\ 3 \times 3,512 \end{array}\right] $ $\times 1$ & $ \left[\begin{array}{c} 1 \times 1,512 \\ 3 \times 3,512 \\ 1 \times 1,2048 \end{array}\right] $ $\times 3$ & $ \left[\begin{array}{c} 1 \times 1,512 \\ 3 \times 3,512 \\ 1 \times 1,2048 \end{array}\right] $ $\times 3$\\[5mm]
        \end{tabular}
        \label{tab:resnet_architecture}
    \end{table}

    Table~\ref{tab:resnet_architecture} summarizes the ResNet-10, ResNet-50, and
    ResNet-101 models. Each architecture is to be read from top to bottom. The
    square brackets indicate a residual or bottleneck block with their two or
    three convolution operations and respective number of filters. The 
    number after the square brackets ``$\times N$'' indicates how often 
    the building block is to be repeated in the respective block group.

    If the output dimension changes from one block group to the next, the first
    building block of the next block group downsamples the dimensions by using a
    stride of 2 for the first 3x3 convolution in the case of a residual block
    and for the middle 3x3 convolution in the case of a bottleneck block. All
    other convolution operations of the building blocks are performed with
    stride 1. 

    We used the ResNet implementation as found in the \emph{TensorFlow} model
    garden \supercite{yuTensorFlowModelGarden}. Since our data is
    one-dimensional, we used an adapted 1D version. We replaced all 2D
    convolutions and pooling operations with their 1D equivalent ($N \times N
    \rightarrow N $). Furthermore, we used a kernel size of 9 in place of the
    3x3 kernels and stride 4 instead of stride 2 in the bottleneck blocks and
    projection skip connections ($N \times N \rightarrow N^2$). This squaring
    was performed to obtain a better distribution of the number of weights
    throughout the architecture (similar to the original 2D case).  
    We further added an additional fully connected layer
    with 256 nodes after the flatten layer in the end of the ResNet models,
    followed by the output layer.

    We were not able to achieve good results when using the original ResNet
    architecture with batch normalization (similar observations were made by \citeauthor{schuetzkeUniversalSyntheticDataset2022a}\supercite{schuetzkeUniversalSyntheticDataset2022a}).
    The test accuracy calculated after
    each epoch was highly unstable and had high fluctuations, probably in part due to
    the moving averages of the batch normalization not converging properly. This
    is possibly caused by using an infinite stream of batches of
    diffractograms, instead of using a training dataset of fixed size. We fixed
    this problem by using group
    normalization\supercite{wuGroupNormalization2018} with 32 groups instead of
    batch normalization.
    
    \subsection{Setup} \label{sec:setup}
    We used a distributed architecture on multiple nodes using the \emph{Python}
    framework \emph{Ray} \supercite{moritzRayDistributedFramework2018a}. The
    configuration utilized throughout this study is visualized in Figure~2b of our paper. 
    Training took place on a \emph{Ray}
    head node with one or two\footnote{The ``parkCNN'' models used a single GPU, while
    the ResNet models were trained on two GPUs.} RTX 2080 Ti GPUs. Training on two
    GPUs was performed using the \emph{MirroredStrategy} of \emph{TensorFlow}
    \supercite{abadiTensorFlowLargeScaleMachine2015}. 28 out of the 32 cores of
    the head node were used for the generation of diffractograms. Next to the
    head node, we used two additional compute nodes with 128 cores each to
    generate diffractograms. Communication between the nodes took place using a
    \emph{Ray queue} object to access the simulated diffractograms.

    Depending on
    the model size and corresponding training speed, this setup allows
    training with up to millions of unique diffractograms per hour.
    For the larger ResNet variants ResNet-50 and ResNet-101, we efficiently used the GPUs. For the ``parkCNN'' and ResNet-10 models, training is fast and the data generation becomes the bottleneck.
    
    To train all models, we used \emph{Keras} \supercite{francoisKeras} with
    \emph{TensorFlow} 2.3 \supercite{abadiTensorFlowLargeScaleMachine2015}.
    Optimization was performed using \emph{Adam}
    \supercite{kingmaAdamMethodStochastic2017} with $ \beta_1=0.9 $ and $
    \beta_2=0.999 $ (\emph{Keras} default parameters). We also tried to use stochastic gradient descent (SGD)
    with momentum and a stepwise learning rate scheduler, but this did not yield
    good results. Depending on the initial conditions, most of the training runs
    using SGD were either unstable or reached low accuracies. For all
    experiments, a cross-entropy loss was utilized.
    
    We used a batch size of $870$ for the ``parkCNN'' models and a batch size
    of $145$ for the three ResNet models. Furthermore, the ``parkCNN'' models
    were trained for $1000$ epochs with a learning rate of $0.001$, while the
    ResNet models were trained for $2000$ epochs with a learning rate of $0.0001$.
    The ResNet models used a step decay of the initial learning rate,
    halving the learning rate after every $500$ epochs.

    When training on diffractograms simulated from synthetic crystals, we used
    $150$ generated batches per epoch when a batch size of $870$ was used
    (``parkCNN medium'' and ``parkCNN big'') and $900$ batches per epoch when a
    batch size of $145$ was used (ResNet). This means that each epoch always
    contained $130\,500$ diffractograms\footnote{Note that the division into epochs
    for training using synthetic crystals is rather arbitrary since each epoch
    contains different diffractograms simulated from different crystals.
    However, the division makes it easy to calculate the performance on the test
    dataset after each epoch.}.

    For the experiments performed directly on diffractograms simulated from ICSD
    crystals, we used the statistics dataset directly to pre-generate the
    training dataset. We excluded the same 85 space groups that were not used by
    the synthetic training due to missing statistics. We used two different
    crystallite sizes for each crystal in the statistics dataset, yielding
    $148\,466 \cdot 2 = 296\,932 $ diffractograms in the training dataset.

    \section{\added{Randomized datasets}}
    \added{In Section~3.3 of the main text, we created randomized variations of the test and statistics datasets to analyze and understand the gap between training on synthetic crystals and testing on the ICSD. For each of the two datasets, we used 22\,500 randomly picked crystals for the analysis. This analysis used the space group labels as reported by the library \emph{PyXtal} (and not the labels as reported by the
    ICSD, which can deviate for a small number of structures). Furthermore,
    \emph{PyXtal} does not support partial occupancies. Therefore, we compared
    the accuracies we obtained on the randomized datasets with a dataset that
    has all occupancies set to $1.0$ and uses the \emph{PyXtal} space group
    labels. We call this the ``reference dataset''. The difference in
    accuracy between this reference dataset and the test / statistics dataset is relatively
    small ($\approx 1$ percentage point).}
    
    \section{Application to experimental diffractograms}
    \subsection{Dataset}
    To test the performance on experimental diffractograms, we used 
    the publicly available RRUFF mineral database \supercite{lafuentePowerDatabasesRRUFF2015}. It contains 5829 mineral samples with multiple types of
    measured spectra and, most important for us, powder XRD measurements with
    K$_\alpha$ radiation for 2952 samples. Of these 2952 samples, 2875 samples
    provide the output of a Rietveld refinement, including the space group label. We further
    removed by hand some samples that had excessive amounts of noise and
    were of bad quality. Many samples further only provide a simulated
    diffractogram without noise or background. They were also excluded. This left
    942 diffractograms for our analysis.

    \subsection{Data generation}
    To be able to apply our models to experimental diffractograms, we
    added an additional background function and noise to the generated
    diffractograms to make them more similar to experimental data. 
    We used a Gaussian process to generate
    random background functions and added additive and multiplicative Gaussian noise.
    All
    diffractograms were generated in the range $ 2 \theta \in \left[ 5, 90
    \right] \si{\degree} $ with step size \SI{0.01}{\degree}.

    The generation protocol is as follows:

    \begin{enumerate}
        \item Sample the background function from a Gaussian
        process with radial basis function kernel without any
        conditioning:
        \begin{equation}
            k(x_i, x_j) = a \exp\left(- \frac{|x_i - x_j|^2}{2l^2} \right)
        \end{equation}
        We chose $ a = 1 $ and sampled $ l $ uniformly in 
        $[7,40]$ for each diffractogram.
        \item Subtract the minimum intensity from the background function obtained from the Gaussian process.
        \item Divide it by the sum of the 8501 ($2\theta$-range) entries.
        \item Multiply it by a scaling factor drawn from a truncated normal
        distribution in the range $[0,150]$ with mean $0$ and standard deviation
        $75$.
        \item Add the pure diffractogram (intensity-range $[0,1.0]$).
        \item Add Gaussian additive noise with mean $0$ and standard deviation uniformly drawn in $[0,0.02]$.
        \item Multiply with multiplicative Gaussian noise with mean $1$ and standard deviation uniformly drawn in $[0,0.16]$.
    \end{enumerate}

    To resemble a more realistic peak profile, we used the pseudo-Voigt profile instead of the Gaussian profile
    that we used for the classification of pure diffractograms.
    The pseudo-Voigt profile uses the full width at half maximum (FWHM) $ \Gamma_\text{G} $ of
    the Gaussian $G$, the FWHM $\Gamma_\text{L}$ of the Lorentzian $L$, and a mixing parameter $\eta$ as parameters \supercite{gilmoreInternationalTablesCrystallography2019}:

    \begin{equation}
        \text{PV}\left(\theta-\theta_{0} ; \Gamma_\text{L}, \Gamma_\text{G} \right)=\eta G\left(\theta-\theta_{0} ; \Gamma_\text{G}\right)+(1-\eta) L\left(\theta-\theta_{0} ; \Gamma_\text{L}\right)
        \label{eq:pseudo_voigt}
    \end{equation}

    with

    \begin{equation}
        G(\theta-\theta_0;\sigma=\frac{\Gamma_\text{G}}{2\sqrt{2 \ln{2}}})=\frac{1}{\sigma \sqrt{2 \pi}} e^{-\frac{1}{2}\left(\frac{\theta-\theta_0}{\sigma}\right)^{2}}
    \end{equation}

    and

    \begin{equation}
        L(\theta-\theta_0;\gamma=\frac{1}{2} \Gamma_\text{L})=\frac{1}{\pi \gamma}\left[\frac{\gamma^{2}}{\left(\theta-\theta_{0}\right)^{2}+\gamma^{2}}\right] \, \text{.}
    \end{equation}
    
    The FWHM $\Gamma_\text{G}$ of the Gaussian is typically 
    parametrized using the Caglioti equation \supercite{gilmoreInternationalTablesCrystallography2019} as
    
    \begin{equation}
        \Gamma_\text{G}^{2}=U \tan ^{2} \theta+V \tan \theta+W \, \text{.}
        \label{eq:caglioti}
    \end{equation}
    
    Following the suggestions for typical Rietveld parameter ranges by
    \citeauthor{kadukTypicalValuesRietveld2011} \supercite{kadukTypicalValuesRietveld2011}
    and comparing the resulting peaks with the ones from the RRUFF dataset, we decided to sample
    the Caglioti parameters uniformly in the following ranges: $ [0,0.01] $ for
    U, $ [0,0.01] $ for W, and V was fixed at $V=0.0$. We further used a single
    mixing parameter $ \eta $ uniformly sampled in $ [0.0, 1.0] $ for the full
    $2\theta$-range. For simplicity, we used
    the same FWHM from the Caglioti equation for both the Gaussian and
    Lorentzian of the pseudo-Voigt. We further considered the K$_{\alpha_1}$ /
    K$_{\alpha_2}$ splitting of the K$_\alpha$ line since for some
    diffractograms in the RRUFF dataset, this splitting is visible.

    We further implemented the option to add impurity phases to the training
    (and simulated ICSD test) data. The minerals of the RRUFF database are not all made up
    of one phase, but most of them contain small amounts of one or more
    impurity phases. To model this, for each training diffractogram, we used a
    superposition of the main phase to be classified and an impurity phase of a
    random space group ($a$ is uniformly sampled in $[0,0.05]$):

    \begin{equation}
        I(\theta) = (1-a) I_\text{pure} + a I_\text{impurity}
    \end{equation}

    \subsection{Experiments}
    For our experiments on experimental data, we used the same
    split based on structure types as we used for pure diffractograms.
    We performed two experiments using the ResNet-50 architecture, one with impurity phases 
    and one without. For both, a learning rate of $0.0001$,
    a batch size of 145, and 1000 epochs where used.
    
    \begin{figure}[!htb] 
    \centering
    \includegraphics{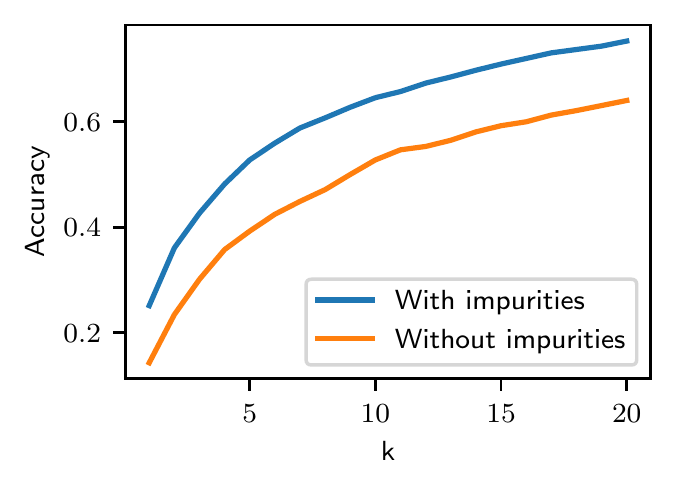}
    \caption{Top-$k$ accuracy as a function of $k$ tested on RRUFF dataset for ResNet-50 model trained with added noise and
    background. The experiment corresponding to the blue curve additionally contained one added impurity phase for
    each diffractogram in the training data.}
    \label{fig:rruff_top_k}
    \end{figure}

\newpage
\section{Additional figures\added{ and tables}}

\begin{figure}[!htb] 
\centering
\includegraphics{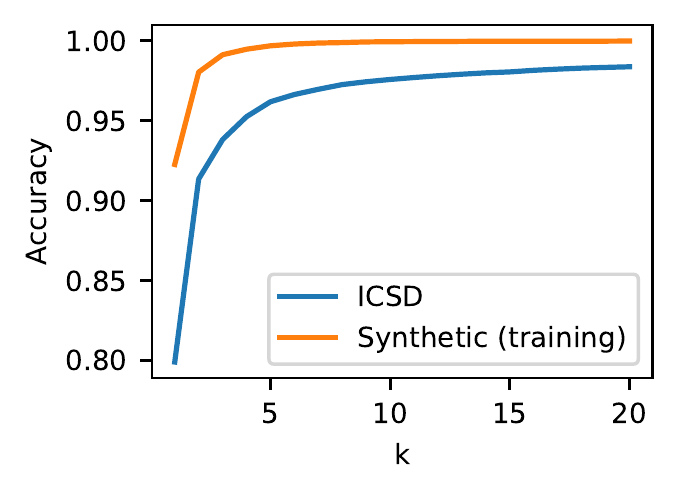}
\caption{Top-$k$ accuracy as a function of $k$ tested on ICSD test dataset and the synthetic training data for ResNet-101 model trained on synthetic data.}
\label{fig:resnet_101_top_k}
\end{figure}
 
\begin{figure}[!htb] 
\centering
\includegraphics{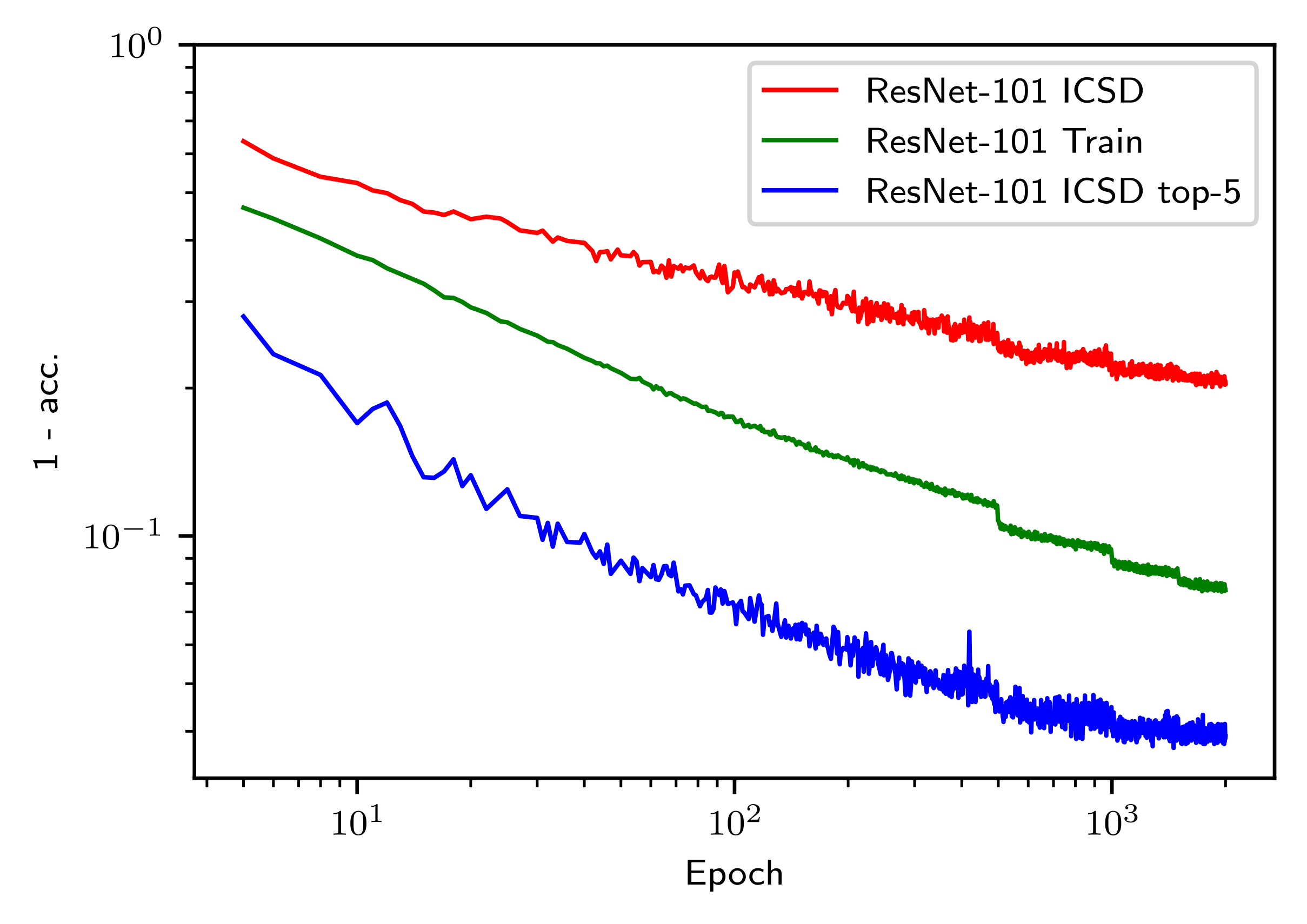}
\caption{$1-\text{acc.}$ for test accuracy (ICSD), training accuracy (synthetic crystals), and
test top-5 accuracy (ICSD) as a function of epochs for ResNet-101 model trained on synthetic data. To better show the scaling
behavior, both axes use logarithmic scaling.}
\label{fig:resnet_101_1_minus_acc}
\end{figure}

\begin{figure}[!htb] 
\centering
\includegraphics{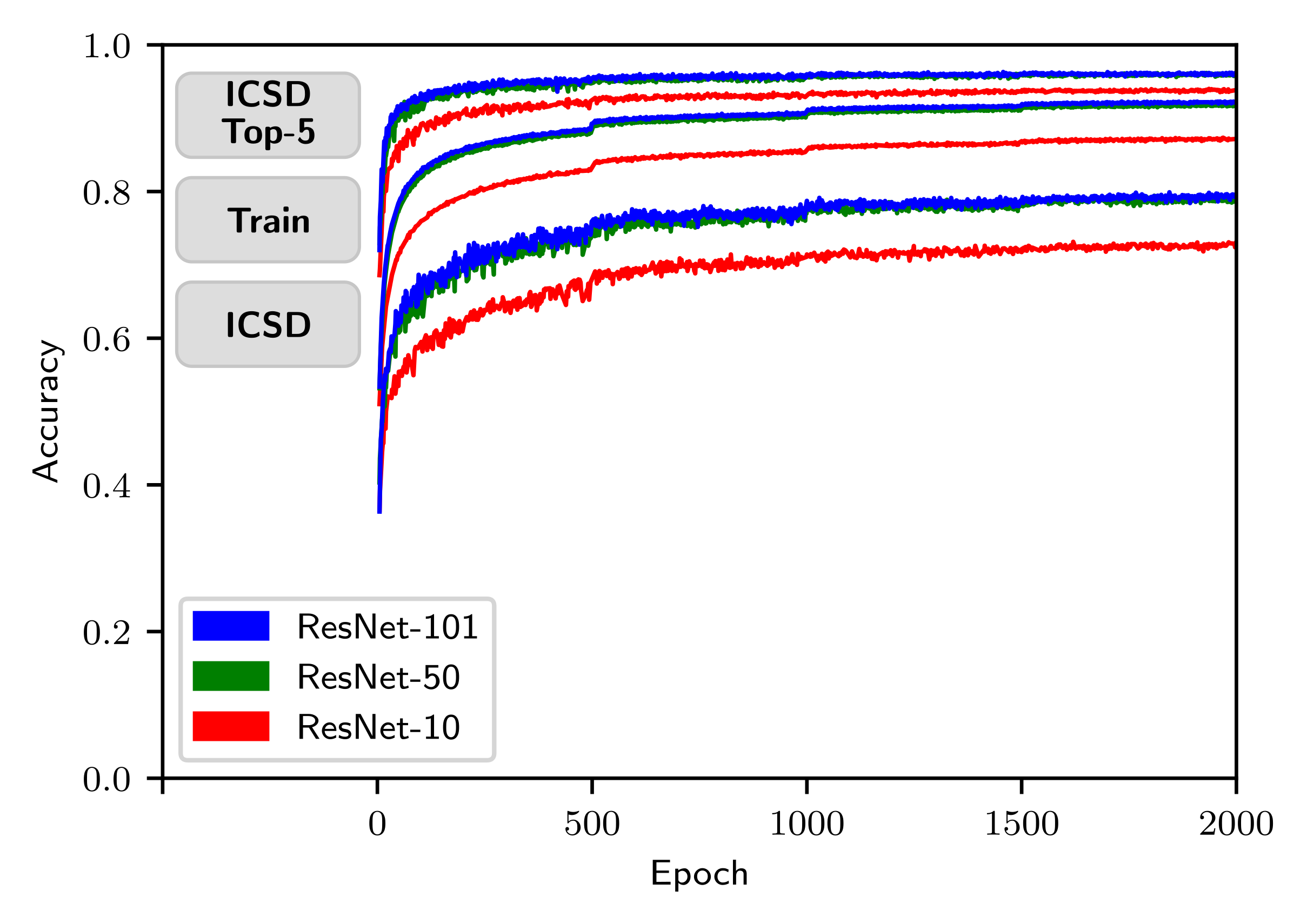}
\caption{\added{Test accuracy (ICSD), training accuracy (synthetic crystals), and
test top-5 accuracy (ICSD) as a function of epochs. We show all three metrics for the
models ResNet-101, ResNet-50, and ResNet-10.}}
\label{fig:training_curve_nolog}
\end{figure}

\begin{table}[!htb]
    \begin{center}
    \caption{\added{This is an extension of Table 1 of the main text.
    We additionally provide the total number of unique diffractograms 
    seen during training and the training time
    for each computational experiment. To obtain the total number of unique diffractograms, we also counted diffractograms that are based on the same crystal structure but have a different crystallite size. To get the number of unique crystals, the provided number for all experiments directly trained on ICSD data and for the experiment using synthetic data with the ``parkCNN big'' model needs to be divided by two, since each crystal is used to generate two diffractograms with different crystallite sizes in those experiments (see Section 2.5 of the main text).
    Training times are based on the hardware setup described in Section \ref{sec:setup}.}}
    \vspace*{2mm}
    
    \begin{tabular}{ccccccc} 
        \toprule
        Split & \begin{tabular}{@{}c@{}}Training \\ dataset\end{tabular} & \begin{tabular}{@{}c@{}}Testing \\ dataset\end{tabular} & Model & \begin{tabular}{@{}c@{}}Total number of \\ unique diffractograms\end{tabular} & \begin{tabular}{@{}c@{}}Training \\ time\end{tabular} \\
        \midrule
        Random & ICSD & ICSD & parkCNN medium & 296\,932 & $\approx$ 1 day \\
        \midrule
        Structure& \multirow{2}{*}{ICSD} & \multirow{2}{*}{ICSD}& parkCNN big & 296\,932 & $\approx$ 1 day\\
        type&&& parkCNN medium & 296\,932 & $\approx$ 1 day \\
        \midrule
        & \multirow{4}{*}{synthetic} & \multirow{4}{*}{ICSD} &parkCNN big & 130\,500\,000 & $\approx$ 1 day\\
        Structure&&& ResNet-10 & 261\,000\,000 & $\approx$ 3.5 days \\
        type$^{1}$ &&& ResNet-50 & 261\,000\,000 & $\approx$ 11 days \\
        &&& ResNet-101 & 261\,000\,000 & $\approx$ 17.5 days \\
        \bottomrule
    \end{tabular}\label{tab:results_extended}
    \end{center} 
    \footnotesize{$^{1}$Here, the split type refers to the statistics and the test dataset, rather than the training and the test dataset.}
\end{table}

\begin{figure}[!htb] 
\centering
\includegraphics{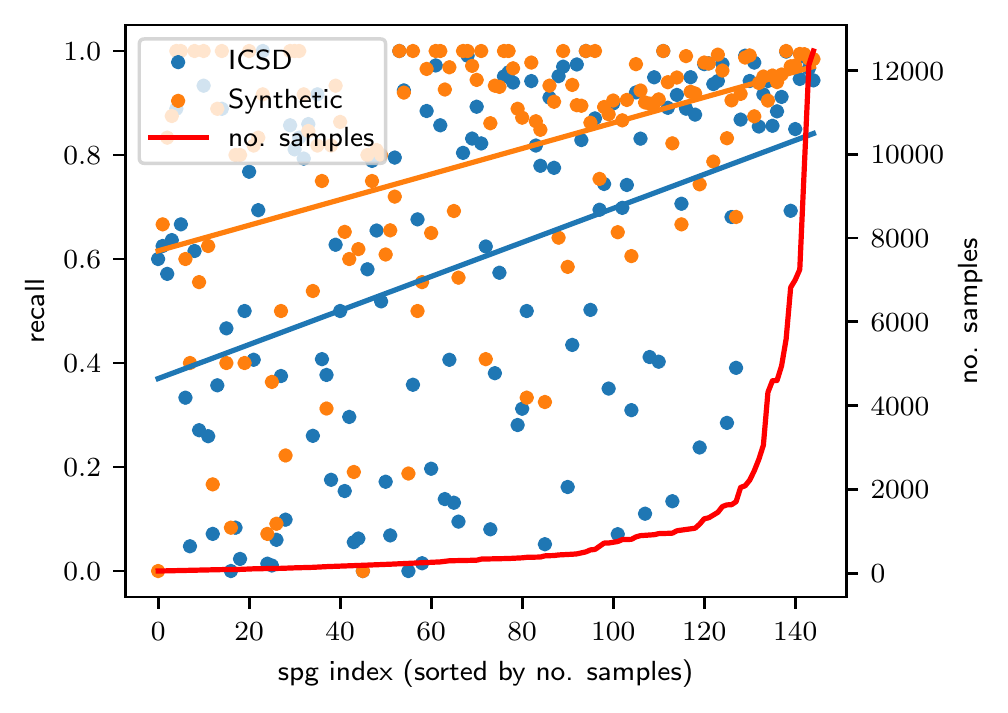}
\caption{\added{Using the ResNet-101 model trained on the synthetic crystals, this figure shows the recall ($\frac{TP}{TP+FN}$) and no. samples for all 145 space groups included in our experiments. Space group numbers are sorted by the no. samples. In blue, one can find the recall tested on the ICSD test dataset, in orange tested on diffractograms from the synthetic training distribution. The blue and orange lines show the trend of the recall.}}
\label{fig:recall_analysis_synthetic}
\end{figure}

\begin{figure}[!htb] 
\centering
\includegraphics{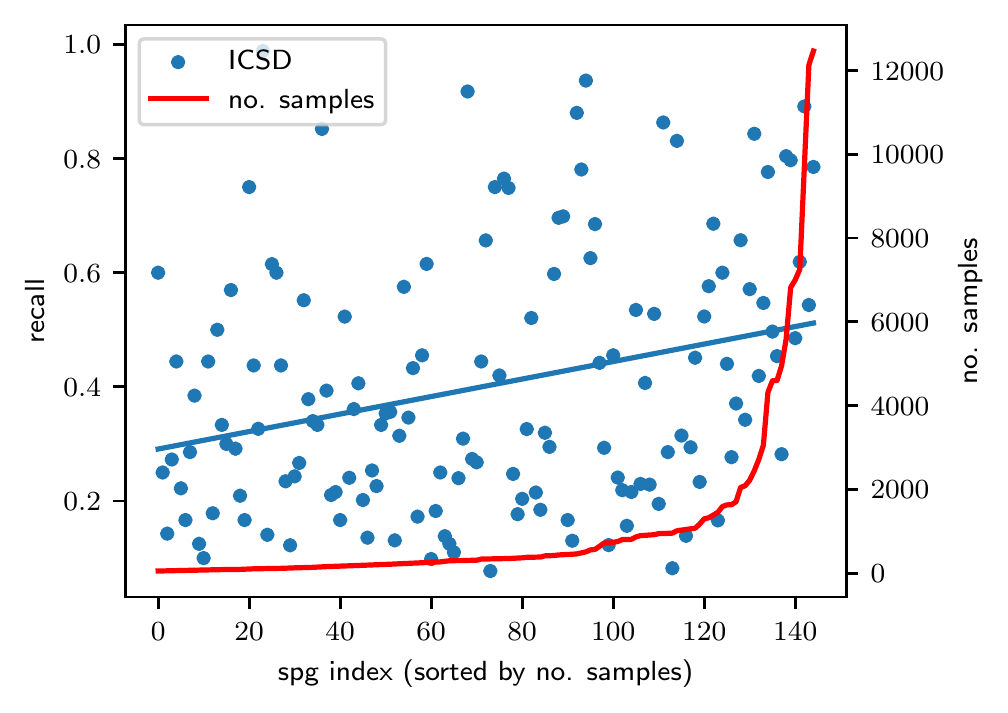}
\caption{\added{Using the ``parkCNN big'' model trained on the ICSD crystals directly, this figure shows the recall ($\frac{TP}{TP+FN}$) and no. samples for all 145 space groups included in our experiments. Space group numbers are sorted by the no. samples. The recall is tested on the ICSD test dataset. The blue line shows the trend of the recall.}}
\label{fig:recall_analysis_direct}
\end{figure}

\begin{figure}[!htb] 
\centering
\includegraphics{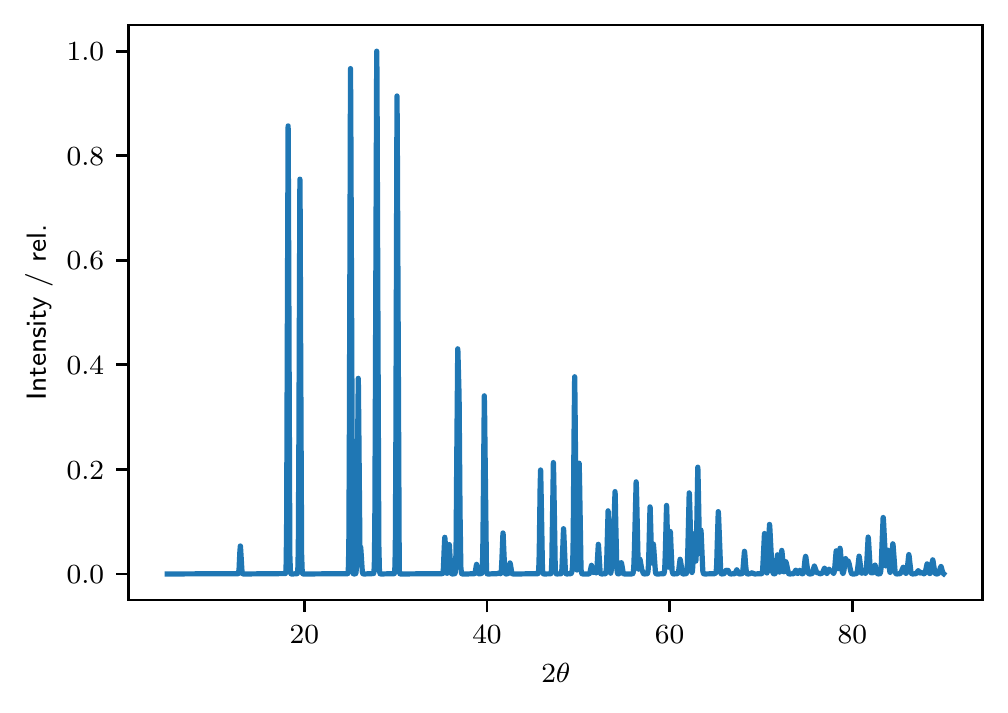}
\caption{\added{Exemplary diffractogram from the ICSD, simulated without imperfections (noise and background).}}
\label{fig:example_diff_ICSD_perfect}
\end{figure}

\begin{figure}[!htb] 
\centering
\includegraphics{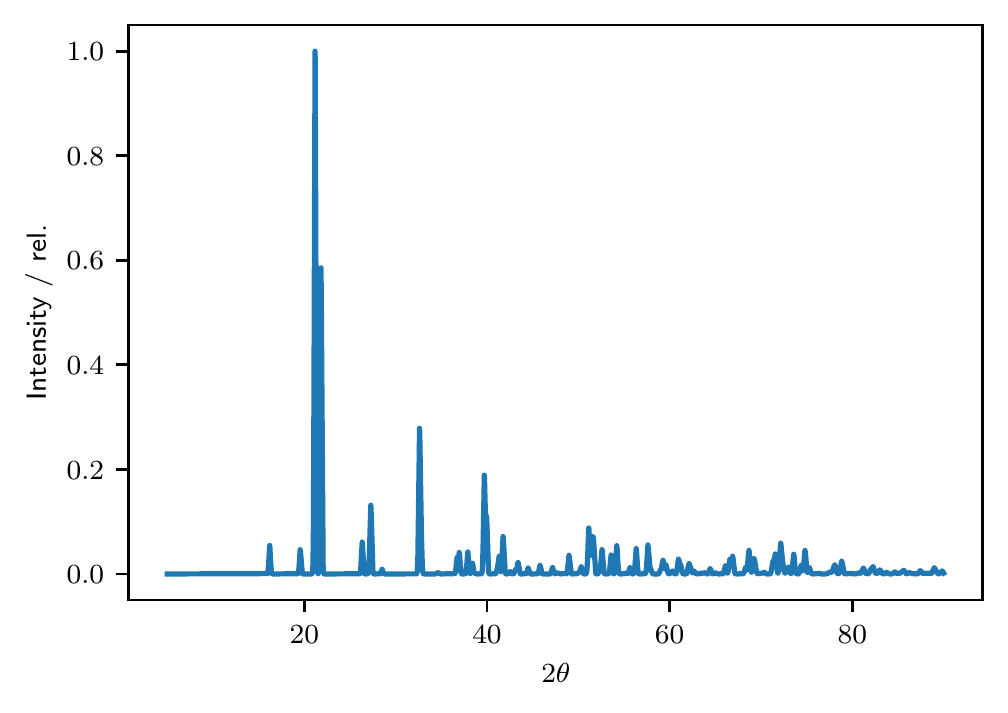}
\caption{\added{Exemplary diffractogram from the synthetic crystal distribution, simulated without imperfections (noise and background).}}
\label{fig:example_diff_synthetic_perfect}
\end{figure}

\begin{figure}[!htb] 
\centering
\includegraphics{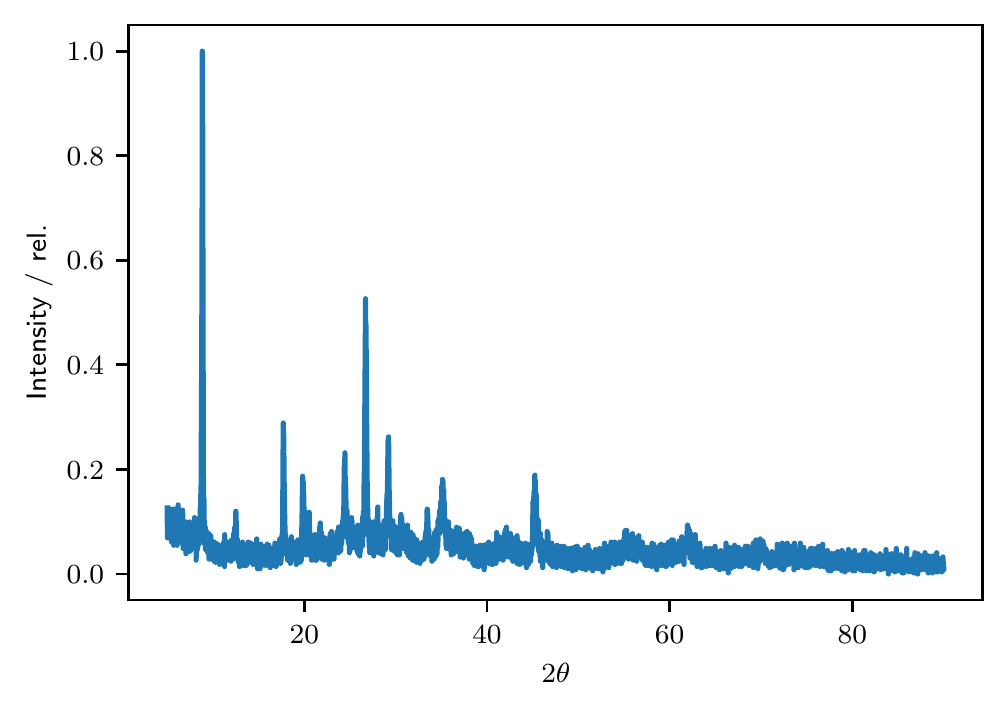}
\caption{\added{Exemplary diffractogram from experimental RRUFF mineral database.}}
\label{fig:example_diff_rruff}
\end{figure}

\begin{figure}[!htb] 
\centering
\includegraphics{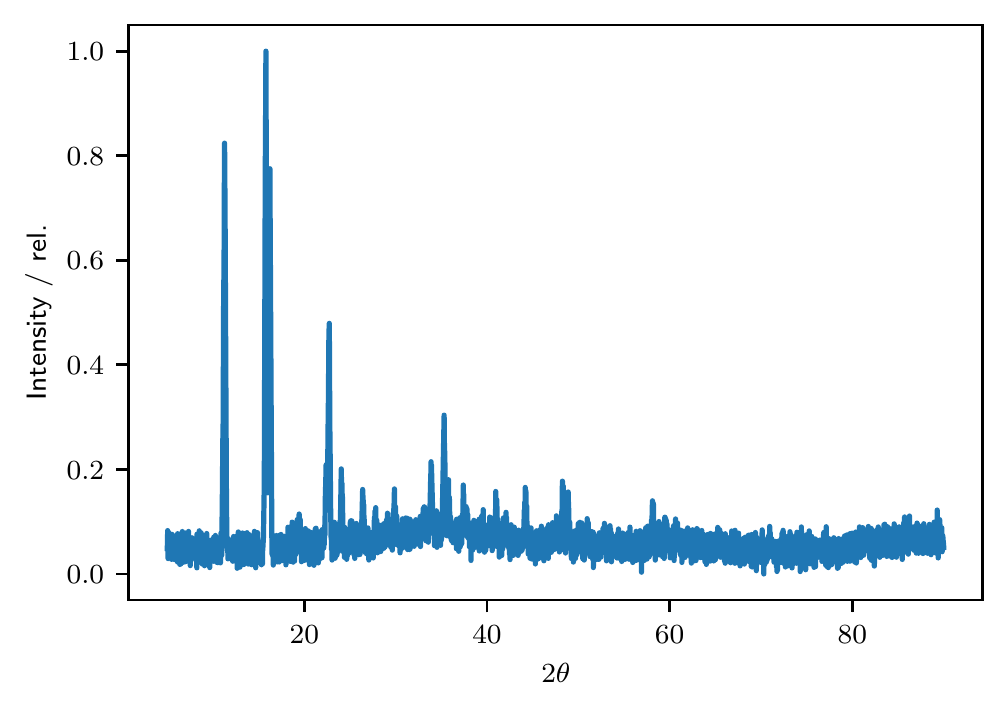}
\caption{\added{Exemplary diffractogram from the synthetic crystal distribution, simulated with imperfections (noise, background, and impurity phase).}}
\label{fig:example_diff_synthetic_imperfect}
\end{figure}

\begin{figure}[!htb] 
\centering
\includegraphics{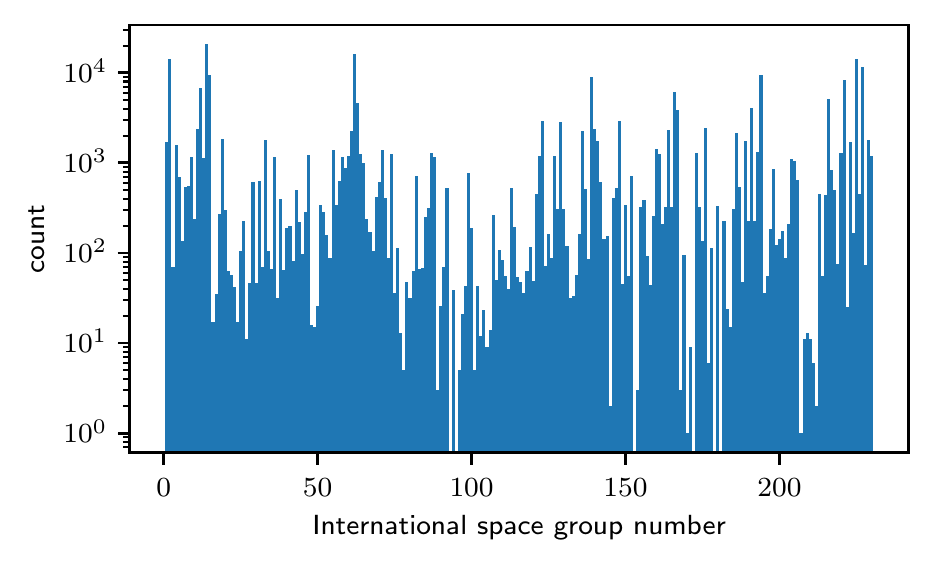}
\caption{\added{Distribution (logarithmic scale) of space groups in the ICSD.}}
\label{fig:distribution_spgs_nolog}
\end{figure}

\printnomenclature

\clearpage

\TheBibliography
\printbibliography[heading=bibintoc]